\documentclass{article}
\usepackage[a4paper,top=2cm,bottom=2cm,left=2cm,right=2cm]{geometry}
\usepackage{authblk}
  \usepackage{paralist}
  \usepackage{graphics}
  \usepackage{epsfig} 
\usepackage{graphicx} 
\usepackage{epstopdf}
 \usepackage[colorlinks=true]{hyperref}
\hypersetup{urlcolor=blue, citecolor=red}
\usepackage{amsmath} 
\usepackage{amsfonts}
\usepackage[bbgreekl]{mathbbol}
\DeclareSymbolFontAlphabet{\mathbbm}{bbold}
\DeclareSymbolFontAlphabet{\mathbb}{AMSb}
\usepackage{amssymb}
\usepackage{amsthm}
\usepackage{mathrsfs}
\usepackage{stackrel,graphicx}
\usepackage{enumitem}
\usepackage{cases}
\usepackage{booktabs}
\usepackage{braket}
\usepackage[strict]{changepage}
\usepackage{subfigure}
\usepackage[small,justification=justified]{caption}
\usepackage{titlesec} 
\usepackage{multicol}
\usepackage{hyperref}
\usepackage{chemarr}

\usepackage{amssymb,amsmath,amsthm}
\usepackage{booktabs}
\usepackage{multirow}
\usepackage[table]{xcolor}
\usepackage{subfigure}
\usepackage{chemfig}
\usepackage[version=4]{mhchem}
\usepackage[right]{lineno}

\newtheorem{oss*}{Observation}
\newtheorem*{rem*}{Remark}
\theoremstyle{definition}

\newcommand{\z}{{\bf z}}

\newcommand{\bZ}{{\bf Z}}

\providecommand{\keywords}[1]{\textbf{\textit{Keywords --}} #1}

\newenvironment{sistem}
{\left\lbrace\begin{array}{@{}l@{}}}
{\end{array}\right.}

\begin{document}

\title{Kinetic modeling of knowledge and wealth dynamics in national and global markets}
\author[1]{Marzia Bisi}
\author[2]{Martina Conte}
\author[1]{Maria Groppi\footnote{Corresponding author: maria.groppi@unipr.it}}
\affil[1]{\centerline {\small Department of Mathematical, Physical and Computer Sciences, University of Parma} \newline \centerline{\small  Parco Area delle Scienze 53/A, 43124, Parma, Italy}}
\affil[2]{\centerline{\small Department of Mathematical Sciences ``G. L. Lagrange'', Politecnico di Torino} \newline \centerline{\small Corso Duca degli Abruzzi 24, 10129, Torino, Italy}} 

\setcounter{Maxaffil}{0}
\renewcommand\Affilfont{\itshape\small}

\maketitle

\begin{abstract}
We propose a kinetic model to describe the dynamical evolution of wealth and knowledge in national and global markets, starting from a microscopic description of individual interactions. The model is built upon interaction rules that account for a strong interdependence between the microscopic variables, influencing agents’ trading and saving propensities, knowledge acquisition, and the stochastic market effects. We begin with a domestic market scenario and extend the framework to international trade, incorporating the possibility of individual transfers between different countries. The dynamics of the system are described through Boltzmann-type equations, which allow for a detailed study of the evolution of the agent distribution in each country. In this context, we study the evolution of macroscopic quantities of the system, focusing on the number density of individuals, and the mean wealth and knowledge of each population, and we discuss these results in relation to existing models in the literature. Finally, under a quasi-invariant trading limit, we derive simplified Fokker–Planck type equations that reveal some emergent behaviors of the system, including the formation of Pareto tails in the long-term wealth and knowledge distributions.
\end{abstract}

\keywords{Boltzmann equation; Wealth and knowledge distribution; Continuous trading limit; Fokker–Planck equations}


\section{Introduction}\label{sec:intro}
Over recent decades, various mathematical approaches, inspired by statistical physics and, in particular, by the Boltzmann equation, have been developed to describe the collective behavior of living systems.\cite{bellomo2025,bellomo2022what} This has led, in turn, to the formulation of kinetic models for many socio-economic problems. Among these, we recall the approaches describing the evolution of wealth distribution,\cite{cordier2005kinetic,pareschi2013interacting} opinion formation,\cite{toscani2006kinetic,ToscaniTosinZanella} the dynamics of crowds or vehicular traffic,\cite{BellomoGibelli,FregugliaTosin} the motion of swarms,\cite{BellomoHa,bellomo2020swarms} birth and death processes,\cite{Greenman} tumor growth,\cite{ConteDzierma,ConteGroppiSpiga,ConteTravaglini} the spread of epidemics and other diseases.\cite{BertagliaBondesanetal,BertagliaPareschi,BisiLorenzani,DellaMarca} 
In these contexts, the agents/animals/cells are characterized by a specific microscopic variable (e.g., wealth, opinion, or viral load), usually referred to as {\it activity},\cite{carrillo2024book} and are assumed  to be subject to binary interactions. To accurately describe individual or cellular behavior, the microscopic interaction rules must be sufficiently detailed to incorporate non-deterministic effects and, possibly, mutual dependencies among microscopic features. 
Furthermore, multiple populations are often involved in socio--economic phenomena, thus giving rise to a system of coupled Boltzmann equations for single--species distribution functions. A unified kinetic description for multi--population systems with multiple microscopic states may be found in Ref.~\cite{bisi2024kinetic}. Analytical results on the long--time behavior of these systems are usually obtained in proper asymptotic quasi--invariant regimes, where integro--differential Boltzmann equations may be approximated by simpler PDEs of Fokker--Planck type. Suitable numerical schemes for Boltzmann or Fokker--Planck operators allow the investigation of situations far from these regimes. Comprehensive reviews of the fundamental tools and recent advances in kinetic models for living systems can be found in Refs.~\cite{bellomo2021,bellomo2025,DiMarcoPareschiZanella,pareschi2013interacting}.

This work is framed within the context of behavioral economy. It aims to model and investigate the combined effects of knowledge and wealth on the dynamics of national and global markets, starting from microscopic descriptions. A fundamental paper on the evolution of a simple market economy is Ref.~\cite{cordier2005kinetic}, where the authors account, besides deterministic trades, possible random effects in the market, which may amplify gains or losses. The presence of random variables allows one to reproduce, at least in mathematically treatable regimes, steady wealth distributions with Pareto tails (i.e., asymptotic to an inverse power law for high values of wealth),\cite{Pareto1897} in agreement with real data. Such kinetic model has then been generalized to allow the presence of some debts,\cite{TorregrossaToscani} taxation and redistribution of collected wealth,\cite{BisiSpigaToscani} and welfare dynamics.\cite{dolfin2014modeling} Further refinements involve non-constant collisional kernels designed to exclude unphysical interactions between agents.\cite{furioli2020nonmax} Moreover, this modeling approach has been extended to a multi--population system, also allowing possible transfers of individuals.\cite{bisi2022kinetic} 

Among the many behavioral aspects influencing national and international trades, this paper focuses on the mutual influence of wealth and knowledge in the evolution of a group of interacting countries.
Knowledge can be understood as familiarity with information, facts, descriptions, or skills acquired through experience or education. It is widely recognized that knowledge is shaped both by familiar influence and by the broader environment in which individuals are raised, including the opportunities to which they have access. The complex interactions between collective knowledge and social dynamics have been matter of various mathematical studies in recent years.\cite{Burini-DeLillo-Symmetry,Liao-etal} A first Boltzmann approach for the description of wealth distribution and collective knowledge in a single population may be found in Ref.~\cite{pareschi2014wealth}. In the present work, we seek to generalize this framework to encompass situations characterized by strong coupling among microscopic variables, within the context of an open market permitting both national and international trade.  The first key novelty of our approach lies in the explicit bidirectional coupling between wealth and knowledge: the evolution of an agent’s knowledge depends on its wealth, while the dynamics of wealth are simultaneously influenced by the agent’s knowledge. Unlike Ref.~\cite{pareschi2014wealth}, which considered only the effect of knowledge on wealth, our model captures this two-way interaction. This reflects more realistic economic scenarios in which individuals with greater wealth have more opportunities to enhance their knowledge, while a solid understanding (knowledge) of market risks can, in turn, reduce unprofitable trades. The second major novelty concerns the introduction of a random variable in the microscopic trade rule, modulated by the agent’s level of knowledge, whose mean value is positive. This feature enables the model to reproduce open-market conditions, where higher knowledge levels among agents can lead to the increase of both personal and global wealth. This is in agreement with historical observations, showing a continuous (slow) increase of world wealth over time. The analysis presented in this paper is mainly devoted to analytically characterizing these phenomena and their emerging properties, as a foundation for future extensions involving numerical and computational investigations.

The paper is organized as follows. Section 2 is devoted to the kinetic modeling of the interplay between wealth distribution and knowledge within a single population. Specifically, mutually dependent interaction rules for the microscopic variables are proposed, and the parallelism between the Boltzmann operator in the classical collisional form and its probabilistic formulation with a transition probability in the kernel is discussed. Then, a suitable quasi--invariant regime is investigated, leading to a Fokker--Planck--type kinetic equation with partial derivatives in both the wealth and knowledge variables.
In Section 3, this way of modeling is extended to a multi--population system, which  describes international trade and allows transfers of individuals from a country to another. For a two--country economy, evolution equations for number density, mean wealth and mean knowledge of each country are derived from the Boltzmann system, and admissible steady states or long--time behavior are commented on. The asymptotic limit leading to a Fokker--Planck system is also extended to the two--population setting, and the relevant steady distributions are discussed in a proper fast--transfer regime. Finally,  Section 4 presents some conclusions and perspectives.

\section{Boltzmann equations for wealth and knowledge evolution in national markets}
In this section, we develop a kinetic model for the joint evolution of wealth and knowledge within a national market, namely in a single population of agents. Domestic trade is modeled through binary, symmetric interaction rules inspired by the classical approaches (see Refs.~\cite{cordier2005kinetic,pareschi2014wealth}), but further effects of mutual influence of personal wealth and knowledge are taken into account.

Each agent is described by a microscopic state $\z=(x,v)\in\mathbb{R}^2_+$, where $x\in\mathbb{R}_+$ represents the individual's knowledge level and $v\in\mathbb{R}_+$ denotes the personal wealth. The wealth dynamics arise from binary interactions between pairs of agents in the population, while the knowledge dynamics result from interactions with a fixed background.
Starting from this microscopic description, we derive the corresponding Boltzmann equations for the joint evolution of wealth and knowledge. We then investigate their structural properties, obtain macroscopic equations for aggregated quantities of interest, and study relevant asymptotic limits in which the integro--differential Boltzmann description reduces to a simpler Fokker–Planck type equation.

\subsection{The Boltzmann model for knowledge and wealth distributions}\label{national_trade}
Following the framework introduced in Ref.~\cite{pareschi2014wealth}, we describe the evolution of knowledge within a population of agents through microscopic interactions with a fixed background. Each knowledge update is interpreted as an interaction in which individuals lose a fraction of their knowledge due to selection mechanisms, while simultaneously acquire new knowledge from the external background, which represents not only the partner interacting agent but the whole surrounding environment. Furthermore, personal wealth plays a central role in shaping knowledge evolution. It is reasonable to assume that individuals with higher wealth levels have greater access to opportunities for acquiring knowledge, thereby reinforcing the mutual dependence between wealth and knowledge dynamics.

The updated level of knowledge of an agent with microscopic state $\z=(x,v)$ can be expressed as
\begin{equation}\label{micro_know}
    x'=(1-\lambda(x))x+\lambda_b(x)b+\beta(v)x+\kappa x\,.
\end{equation}
Here, the function $\lambda(x)$ represents the fraction of knowledge lost due to the natural selection mechanism (each individual forgets a fraction of information). On the other hand, $\lambda_b(x)$ represents the fraction of knowledge gained from the external learning, i.e., from interactions with the background. Indeed $b\ge0$ denotes the maximum amount of knowledge gained from the background during a single interaction. The probability distribution $C(b)$ of background knowledge is assumed normalized and with bounded mean: 

$$\int_{\mathbb{R}_+}C(b)db=1\qquad \text{and} \qquad \int_{\mathbb{R}_+}bC(b)db=\bar{B}\,.$$
The non--negative function $\beta(v)$ in \eqref{micro_know} models the influence of personal wealth on knowledge acquisition, while $\kappa$ is a random variable accounting for unpredictable fluctuations in the knowledge updating process. In general, we assume that $\kappa$ has zero mean, i.e., $\langle\kappa\rangle=0$, and variance $\langle\kappa^2\rangle=\delta^2$. Here and below angular brackets $\langle \cdot \rangle$ denote the mathematical expectation with respect to the involved random variables. As selection mechanisms are always present, though they cannot reduce knowledge completely, we assume bounds on the selection fraction $\lambda_{-}\le \lambda(x)\le \lambda_{+}$, with $\lambda_{-}>0$ and $\lambda_{+}<1$. Similarly, it is natural to assume that the fractions of knowledge acquired from the background and through personal wealth are bounded, namely $0\le\lambda_b(x)\le\bar{\lambda}_b<1$ and $\beta_-\le\beta(v)\le\beta_+$, with $\beta_-,\beta_+\ge 0$. To guarantee that the post-interaction knowledge $x'$ remains non-negative, the random fluctuation term is assumed satisfying $\kappa\ge -(1-\lambda_{+})-\beta_-$.

The introduction of wealth dependence in the knowledge evolution allows for more realistic scenarios, where knowledge loss due to selection may be compensated even in cases where the interaction with the environment alone would not suffice. For instance, consider the case where the background is uniformly distributed on $(0,\bar{b})$ with $\bar{b}>0$, the selection and background acquisition rates are constant, ${\lambda(x)=\lambda_b(x)=\bar{\lambda}}$, fluctuations vanish, $\kappa=0$, and the individual has initial knowledge $x>\bar{b}$. In such situations $x' = (1 - \bar{\lambda}) x + \bar{\lambda}\,b + \beta(v)\,x$, and the wealth-dependent term $\beta(v)x$ would provide a restorative mechanism against knowledge decay.

Regarding the modeling of wealth distribution, numerous kinetic-type models have been proposed under the assumption of indistinguishable agents, whose state at time $t\ge 0$ is fully determined by their wealth $v\ge0$. \cite{cordier2005kinetic,during2008kinetic,during2009boltzmann,matthes2008steady} In Ref.~\cite{cordier2005kinetic}, the post-trade wealths $v'$ and $w'$ of two interacting agents are governed by linear exchange rules involving random variables with fixed distributions, independent of the pre-trade wealths $v$ and $w$. Extending this framework, Ref.~\cite{matthes2008steady} incorporates the notion that investment decisions are inherently risky, allowing for stochastic gains or losses proportional to an agent’s current wealth. This coupling between saving propensity and risk produces interaction rules that generate a broad spectrum of realistic wealth distributions in multi-agent systems. A further refinement is introduced in Ref.~\cite{pareschi2014wealth}, where both saving propensity and risk are modeled as functions of an agent’s personal knowledge. This reflects the idea that knowledge can be leveraged to reduce exposure to unfavorable outcomes or to increase the likelihood of profitable trades. Based on this latter approach, we introduce the assumption that risk-related fluctuations in the trading process are governed by non-centered random variables with a small but nonzero mean. Specifically, for two agents with states $\z=(x,v)$ and $\z_*=(y,w)$, their interaction is modeled through a symmetric binary trade governed by the following rules:
\begin{equation}\label{micro_wealth}
    \begin{sistem}
        v'=\left(1-\gamma\psi(x)\right)v+\gamma\psi(y)w+\phi(x,v)\eta_1+\mu_1v\\[0.3cm]
        w'=\left(1-\gamma\psi(y)\right)w+\gamma\psi(x)v+\phi(y,w)\eta_2+\mu_2w\,.
    \end{sistem}
\end{equation}
The microscopic interaction rules consist of two components: a deterministic part, which besides the overall saving propensity $\gamma$ captures individual saving behavior through the function $\psi(x)$, and stochastic fluctuations, capturing unpredictable and risk-related events. Following the approach in Ref.~\cite{pareschi2014wealth}, in order to reflect the idea that greater knowledge allows individuals to conduct trades more effectively and mitigate losses, the function $\psi(x)$ should be assumed non--increasing with respect to individual knowledge $x$, but also different options could be investigated. The random terms $\mu_1$ and $\mu_2$ describe unpredictable fluctuations in the wealth exchange. For simplicity, we set $\mu_1=\mu_2=\mu$ where $\mu$ is a centered random variable with mean $\langle\mu\rangle=0$ and variance $\langle\mu^2\rangle=\sigma^2$. In contrast, $\eta_1$ and $\eta_2$ represent risk-related fluctuations, whose intensity depends on both wealth and knowledge through the function $\phi(x,v)$. 
For simplicity, we assume $\eta_1=\eta_2=\eta$, where $\eta$ is a non-centered random variable with mean $\langle\eta\rangle=\bar{\eta} \geq 0$ and variance $\langle\eta^2\rangle=\omega^2$. Both the mean and variance of $\eta$ are assumed to be small, but a positive $\bar{\eta}$ allows to take into account in this model that the total wealth is not conserved across interactions, but it increases. In fact, it is given by:
\begin{equation} \label{total-mean-wealth}
    \langle v'+w'\rangle=\langle v+w\rangle +\bar{\eta}\langle\phi(x,v)+\phi(y,w)\rangle\,.
\end{equation}
Various possible choices for the distribution of $\eta$ and $\mu$ have been discussed in the literature, leading to insightful results on the intrinsic risk dynamics within financial markets.\cite{matthes2008steady} We also remark that other possibilities to reproduce the historically observed growth of global wealth have been investigated by introducing proper indicator functions in the kernels of interaction operators.\cite{cordier2005kinetic}

Assuming that the binary trade rule \eqref{micro_wealth} governs the microscopic interactions of wealth exchange, and that \eqref{micro_know} prescribes the evolution of knowledge, the joint dynamics of wealth and knowledge within the agent system are described by the density function $f=f(t,\z)$. Specifically, 
$f=f(t,\z)$ denotes the distribution of agents at time $t>0$ characterized by the state $\z=(x,v)\in \mathbb{R}_+^2$, with $x$ representing knowledge and $v$ representing wealth. The time evolution of this density is governed by a Boltzmann-type equation, which in weak form reads
\begin{equation}\label{gen_kin_eq}
\begin{split}
    &\dfrac{d}{dt}\int\limits_{\mathbb{R}^2_+}\varphi(\z )f(t,\z)d\z =\\
    &\dfrac{1}{2}\left\langle\,\,\int\limits_{\mathbb{R}^5_+} \chi_{\z\z_*} \left(\varphi(\z')+\varphi(\z'_*)-\varphi(\z)-\varphi(\z_*)\right)C(b)f(t,\z)f(t,\z_*)d\z d\z_*db\,\,\right\rangle\,.
    \end{split}
\end{equation}
In \eqref{gen_kin_eq}, the post-interaction states $\z'$ and $\z_*'$ are obtained from $\z$ and $\z_*$ by \eqref{micro_know} and \eqref{micro_wealth} while, as described above, $\langle\cdot \rangle$ represents mathematical expectation that takes into account the fact that the interaction coefficients involve random variables. Moreover, $\chi_{\z\z_*}$ represents the interaction frequency between the agents with states $\z$ and $\z_*$, which in case of indistinguishable individuals is symmetric, i.e. $\chi_{\z\z_*}=\chi_{\z_*\z}$. For simplicity, from now on we assume a constant interaction probability $\chi_{\z\z_*}= \chi$.

By choosing the test function $\varphi$ independent of wealth, that is $\varphi=\varphi(x)$, equation \eqref{gen_kin_eq} yields the evolution of the marginal density of knowledge, defined by
\begin{equation}\label{marg_know}
    F_x(t,x):=\int_{\mathbb{R}_+}f(t,\z)dv\,.
\end{equation}
Its evolutionary equation (in weak form) reads
\begin{equation}\label{eq_marginal_x}
    \dfrac{d}{dt}\int\limits_{\mathbb{R}_+}\varphi(x)F_x(t,x)dx =\left\langle\chi\int\limits_{\mathbb{R}^2_+}\left(\varphi(x')-\varphi(x)\right)C(b)F_x(t,x)dxdb\right\rangle\,.
\end{equation}
At any positive time $t>0$, this equation describes how the distribution of knowledge evolves due to agent–environment interactions of type \eqref{micro_know}. It is immediate to show that equation \eqref{eq_marginal_x} preserves the total mass, so that, if $F_x(0,x)$ is a probability density, then 
$F_x(t,x)$ remains a probability density for all $t>0$. In particular, choosing $\varphi(x)=x$ provides the dynamics of the mean knowledge
\begin{equation}\label{mean_know}
    M_x(t):=\int_{\mathbb{R}^2_+}xf(t,\z)d\z\,,
\end{equation}
which satisfies
\begin{equation}\label{mean_know_eq}
    \dfrac{d}{dt} M_x(t) =-\chi\int\limits_{\mathbb{R}^2_+}\lambda(x)xf(t,\z)d\z +\chi\bar{B}\int\limits_{\mathbb{R}^2_+}\lambda_b(x)f(t,\z)d\z +\chi\int\limits_{\mathbb{R}^2_+}\beta(v)xf(t,\z)d\z \,.
\end{equation}
Depending on the specific choices of $\lambda(x)$, $\lambda_b(x)$, and $\beta(v)$, one may obtain a closed-form evolution for $M_x(t)$. More generally, under the boundedness assumptions imposed on these functions, the following estimate holds:
\begin{equation}\label{est_mean_know}
    \dfrac{d}{dt} M_x(t)\le \chi(\beta_+-\lambda_{-})M_x(t)+\chi\bar{{\lambda}}_b\bar{B}\,.
\end{equation}
Under the assumption $\lambda_->\beta_+$, which prevents the effect of wealth-dependent acquisition ($\beta(v)$) from overwhelming the dissipative effect of selection ($\lambda(x)$), we note that
 the mean knowledge never exceeds the finite bound
$$
M_x^{max}=\dfrac{\bar{\lambda}_b\bar{B}}{\lambda_--\beta_+}\,.
$$
In the special case where $\lambda(x)=\lambda$, $\lambda_b(x)=\lambda_b$, and $\beta(v)=\beta$ are constants, equation \eqref{mean_know_eq} reduces to
\begin{equation}
    \dfrac{d}{dt} M_x(t)=-\chi(\lambda-\beta)M_x(t)+\chi\lambda_b\bar{B}
\end{equation}
which admits the explicit solution
\begin{equation}\label{Mx_explicit}
M_x(t)=M_{x}(0)e^{-{\chi(\lambda-\beta)}t}+\dfrac{\lambda_b\bar{B}}{(\lambda-\beta)}\left(1-e^{-{\chi(\lambda-\beta)}t}\right)\,,
\end{equation}
showing that, if $\lambda> \beta$, the mean knowledge exponentially converges to the equilibrium value $\lambda_b\bar{B}/(\lambda-\beta)$. Of course for general functions $\lambda(x)$, $\lambda_b(x)$, $\beta(v)$ the boundedness of the mean knowledge is not guaranteed for long times.

Analogously, by choosing the test function $\varphi$ in \eqref{gen_kin_eq} independent of knowledge, i.e., $\varphi=\varphi(v)$, one obtains the evolution of the marginal wealth density, defined as
\begin{equation}\label{marg_wealth}
    F_v(t,v):=\int_{\mathbb{R}_+}f(t,\z)dx\,.
\end{equation}
Its evolutionary equation (in weak form) is given by
\begin{equation}\label{eq_marginal_v}
\begin{split}
    &\dfrac{d}{dt}\int\limits_{\mathbb{R}_+}\varphi(v)F_v(t,v)dv=\\
    &\dfrac{1}{2}\left\langle\,\,\chi\int\limits_{\mathbb{R}^2_+}\left(\varphi(v')+\varphi(w')-\varphi(v)-\varphi(w)\right)F_v(t,v)F_w(t,w)dvdw\,\,\right\rangle\,.
    \end{split}
\end{equation}
At any positive time $t>0$, this equation describes the time evolution of the wealth distribution resulting from binary, symmetric interactions of type \eqref{micro_wealth}. It is immediate to verify that \eqref{eq_marginal_v} conserves the total mass: if $F_v(0,v)$ is a probability density, then $F_v(t,v)$ remains a probability density for all $t>0$. In particular, setting $\varphi(v)=v$ yields the dynamics of the mean wealth
\begin{equation}\label{mean_wealth}
    M_v(t):=\int_{\mathbb{R}^2_+}vf(t,\z)d\z\,,
\end{equation}
which satisfies
\begin{equation}\label{mean_wealth_eq}
    \dfrac{d}{dt} M_v(t)=\chi\bar{\eta}\int\limits_{\mathbb{R}^2_+}\phi(x,v)f(\z,t)d\z \,.
\end{equation}
As expected, since interactions of type \eqref{micro_wealth} do not preserve the total wealth (see \eqref{total-mean-wealth}), the mean wealth $M_v(t)$ is not constant in time. Hence, the only conserved quantity is the total mass. The qualitative behavior of $M_v(t)$ depends on the specific choice of the function $\phi(x,v)$. In general, a closed equation may not be available, but if $\bar{\eta}>0$ and $\phi(x,v)>0$ (at least on a subset of $\mathbb{R}^2_+$ of non--zero measure) then the total (and the mean) wealth is strictly increasing in time. In the simplified case where $\phi(x,v)=\phi(v)=v$, i.e., when personal knowledge does not affect risk-related fluctuations and the perception of risk is linearly increasing in $v$, one obtains
\begin{equation}
M_v(t)=M_{v}(0)e^{\chi\bar{\eta}t}
\end{equation}
which shows that the mean wealth grows exponentially. 

In general, the microscopic interaction rules for knowledge \eqref{micro_know} and trade \eqref{micro_wealth} exhibit a highly intricate structure, since they depend nonlinearly on both microscopic variables. As a consequence, a direct analytical study of the full kinetic equation \eqref{gen_kin_eq} is extremely challenging. Therefore, in the following sections we introduce suitable simplifications of the system in order to make the analysis more tractable.

\subsubsection{A parallelism with the jump process-like models}
Before proceeding with the study of the kinetic system, it is useful to recall that Boltzmann-type models admit a natural probabilistic interpretation in terms of stochastic particle dynamics. In particular, one can establish a correspondence between the microscopic binary interaction rules \eqref{micro_know}–\eqref{micro_wealth}, which underlie the kinetic equation \eqref{gen_kin_eq}, and transition mechanisms described by Markovian jump processes, providing an alternative stochastic formulation of the dynamics. Following the approach in Ref.~\cite{loy2020markov}, we investigate the parallelism between the Boltzmann-type model \eqref{gen_kin_eq} with \eqref{micro_know}–\eqref{micro_wealth} and Markov jump processes governed by transition probabilities. This latter viewpoint connects naturally with the Waldmann probabilistic representation of the Boltzmann equation.\cite{bisi2024kinetic,waldmann1958transporterscheinungen,boffi1990equivalence}

If two agents with microscopic states $\z=(x,v)$ and $\z_*=(y,w)$ undergo a symmetric binary interaction, the post-interaction state $\z'=(x',v')$ is defined by 
\begin{equation}\label{micro_state}
    \begin{sistem}
    x'=\tilde{I}(x,v)+\tilde{D}(x)\kappa,\\[0.3cm]
    v'=\hat{I}(x,v,y,w)+\hat{D}_1(x,v)\eta+\hat{D}_2(v)\mu,
    \end{sistem}
\end{equation}
where the deterministic components are  
\[
\tilde{I}(x,v)=(1-\lambda(x))x+\lambda_b(x)b+\beta(v)x, 
\quad
\hat{I}(x,v,y,w)=\bigl(1-\gamma\psi(x)\bigr)v+\gamma\psi(y)w,
\]  
and the diffusion coefficients, quantifying the stochastic fluctuation intensities, are given by  
\[
\tilde{D}(x)=x, 
\qquad 
\hat{D}_1(x,v)=\phi(x,v), 
\qquad 
\hat{D}_2(v)=v.
\]  
We remark that the interaction is symmetric: exchanging $x$ with $y$ and $v$ with $w$ in \eqref{micro_state} yields the post-interaction state $\z_*'=(y',w')$. Let $f=f(t,\z)$ denote the probability density of agents at time $t>0$ with state $\z$. Then, the equation 
\begin{equation}\label{gen_kin_eq_sym}
    \dfrac{d}{dt}\int\limits_{\mathbb{R}^2_+}\varphi(\z )f(t,\z)d\z =\left\langle\chi\int\limits_{\mathbb{R}^5_+}\left(\varphi(\z')-\varphi(\z )\right) C(b)f(t,\z)f(t,\z_*)d\z d\z_*db\right\rangle
\end{equation}
 provides the kinetic formulation in weak form associated with the symmetric interaction rules \eqref{micro_state}.  

Alternatively, the changes in the microscopic state can be described as the outcome of symmetric stochastic binary interactions encoded by a transition probability function   
\begin{equation}\label{trans_prob}
    T(\z'|\z,\z_*)>0 \qquad \forall \,\z,\z_*\in\mathbb{R}^2_+,
\end{equation}
which is a conditional probability density satisfying  
\[
\int\limits_{\mathbb{R}^2_+} T(\z'|\z,\z_*)\, d\z'=1.
\]  
Here $T(\z'|\z,\z_*)$ denotes the probability that, in a binary symmetric interaction between $\z$ and $\z_*$, the state $\z$ changes into $\z'$. Following Ref.~\cite{pareschi2013interacting}, we may formalize this in terms of a discrete-time stochastic process. Let $\bZ_t,\bZ_{*,t}$ be random variables representing the states of two agents at time $t>0$, and let $f=f(t,\z)$ be the probability density function associated to the multi-agent system, i.e., the probability density function of the random variable of a given agent $\bZ_t$. 
Over a small time interval $\Delta t>0$, the state updates according to  
\begin{equation}\label{micro_random_proc}
    \bZ_{t+\Delta t}=(1-\Sigma)\bZ_t+\Sigma\bZ_t',
\end{equation}
where $\bZ_t'$ is the post-interaction state with joint density ${g=g(\bZ_t'=\z';\bZ_t=\z,\bZ_{*,t}=\z_*)}$, and $\Sigma\in\{0,1\}$ is a Bernoulli random variable (independent of the randomness in \eqref{micro_state}), indicating whether a binary interaction occurs  ($\Sigma=1$) or not ($\Sigma=0$) in the time interval $\Delta t$. It takes value $1$ with probability $\text{Prob}(\Sigma=1)=\chi\Delta t$, where $\chi$ denotes the interaction frequency between the agents. Given an observable quantity $\varphi(\z)$, defined on $\z\in\mathbb{R}^2_+$, its mean variation rate in the interval $\Delta t$ is 
\begin{equation}\label{mean_variation}
   \frac{\langle\varphi(\bZ_{t+\Delta t})\rangle-\langle\varphi(\bZ_{t})\rangle}{\Delta t}  =\frac{\langle(1-\chi\Delta t)\varphi(\bZ_t)\rangle + \langle \chi\Delta t\varphi(\bZ_t')\rangle -\langle\varphi(\bZ_t)\rangle}{\Delta t}.
\end{equation}  
Passing to the limit $\Delta t\to0^+$ yields the instantaneous time variation   
\begin{equation}\label{inst_variation}
\frac{d}{dt}\langle\varphi(\bZ_t)\rangle=\left\langle \chi\bigl(\varphi(\bZ_t')-\varphi(\bZ_t)\bigr)\right\rangle .
\end{equation}  
In order to link this to the kinetic description, note that the joint density is defined as  
\[
g(\z';\z,\z_*)=T(\z'|\z,\z_*) f_2(\z,\z_*,t),
\]  
where $f_2$ is the two-particle joint density. Under the propagation of chaos assumption, equation \eqref{mean_variation} then becomes   
\begin{equation}\label{kin_eq_transProb}
    \frac{d}{dt}\int\limits_{\mathbb{R}^2_+} f(t,\z)\varphi(\z)\,d\z
    =\chi\int\limits_{\mathbb{R}^2_+}\int\limits_{\mathbb{R}^5_+}\, T(\z'|\z,\z_*)(\varphi(\z')-\varphi(\z)) f(t,\z)f(t,\z_*)C(b)\, db\, d\z\, d\z_*\, d\z' .
\end{equation}  
Recalling the first order moments of $f$, namely $M_x(t)$ and $M_v(t)$ defined in \eqref{mean_know} and \eqref{mean_wealth}, respectively, and introducing its higher order moments  
\[
E_x(t)=\int\limits_{\mathbb{R}^2_+}x^2 f(t,\z)\,d\z,\quad
e_x(t)=\int\limits_{\mathbb{R}^2_+}(x-M_x)^2 f(t,\z)\,d\z,
\]  
\[
E_v(t)=\int\limits_{\mathbb{R}^2_+}v^2 f(t,\z)\,d\z,\quad
e_v(t)=\int\limits_{\mathbb{R}^2_+}(v-M_v)^2 f(t,\z)\,d\z,
\]
we compare their evolution under \eqref{micro_state}–\eqref{gen_kin_eq_sym} and under \eqref{micro_random_proc}-\eqref{kin_eq_transProb}. Choosing $\varphi(\z)=x,x^2,v,v^2$ in both formulations yields identical dynamics for the mean provided that
\begin{equation}\label{Vtx_Vtv}
V^T_x(\z,\z_*)=\tilde{I}(x,v),\qquad 
V^T_v(\z,\z_*)=\hat{I}(x,v,y,w)+\bar{\eta}\hat{D}_1(x,v),
\end{equation}  
where  
\[
V^T_x(\z,\z_*):=\int\limits_{\mathbb{R}^2_+} x' T(\z'|\z,\z_*)\, d\z', 
\qquad
V^T_v(\z,\z_*):=\int\limits_{\mathbb{R}^2_+} v' T(\z'|\z,\z_*)\, d\z',
\] 
are the mean values of $T$ with respect to $x$ and $v$, respectively. From \eqref{Vtx_Vtv}, we see that the mean of $T$ with respect to $x$ depends only on $\z$, i.e., ${V^T_x=V^T_x(\z)=V^T_x(x,v)}$. Similarly, comparing the second order moments yields  
\begin{equation}\label{DT_x}
\tilde{D}(x)=\frac{1}{\delta}\sqrt{E^T_x(\z,\z_*)-(V^T_x(\z))^2}=:D_x^T
\end{equation}   
with  
\[
E^T_x(\z,\z_*)=\int\limits_{\mathbb{R}^2_+} (x')^2 T(\z'|\z,\z_*)\, d\z'\,
\]
the second order moment of $T$ with respect to $x$. Moreover we have  
\begin{equation}\label{DT_v}
(\hat{D}_2)^2(v)+\frac{\omega^2-\bar{\eta}^2}{\sigma^2}(\hat{D}_1)^2(x,v)
=\frac{1}{\sigma^2}\Bigl(E^T_v(\z,\z_*)-(V^T_v(\z,\z_*))^2\Bigr)=:(D_v^T)^2
\end{equation} 
with  
\[
E^T_v(\z,\z_*)=\int\limits_{\mathbb{R}^2_+} (v')^2 T(\z'|\z,\z_*)\, d\z',
\] 
the second order moments of $T$ with respect to $v$. In particular, from \eqref{DT_x} and \eqref{DT_v} it follows that $D_x^T=D_x^T(x)$ and $D_v^T=D_v^T(x,v)$. Thus, in terms of transition probabilities, we can define an equivalent interaction model of the form  
\begin{equation}\label{micro_state_parallel}
    \begin{sistem}
    x' = V^T_x(x,v) + D^T_x(x)\kappa,\\[0.3cm]
    v' = V^T_v(x,v,y,w) + D^T_v(x,v)\mu,
    \end{sistem}
\end{equation}
which ensures the equivalence between \eqref{kin_eq_transProb} and \eqref{gen_kin_eq_sym} at the macroscopic level, at least for the first- and second-order moments.

We emphasize that, given a collisional model of the form \eqref{micro_state}, it is always possible to construct a transition probability $T$ such that the corresponding jump-process formulation \eqref{kin_eq_transProb} reproduces the same macroscopic evolution of the statistical moments. Conversely, starting from a transition probability $T$, one may define a collisional model of type \eqref{micro_state_parallel} whose corresponding kinetic formulation \eqref{gen_kin_eq_sym} preserves the same first- and second-order moments. However, the kinetic equation \eqref{gen_kin_eq_sym} with interaction rules \eqref{micro_state} is not equivalent to the kinetic formulation \eqref{kin_eq_transProb}.
Indeed, even under the strong assumption
\begin{equation}
T(\z'|\z,\z_*)=\delta\left(x'-(V^T_x(x,v) + D^T_x(x)\kappa)\right)\delta\left(v'-(V^T_v(x,v,y,w) + D^T_v(x,v)\mu)\right)
\end{equation}
where $\delta$ denotes the Dirac delta, one cannot formally recover the wealth dynamics defined by the original collisional model \eqref{micro_state}. The reason is that the knowledge-dependent risk fluctuation (described by $\eta$) and the random market fluctuation (described by $\mu$) are coupled into a single term, rather than appearing as distinct contributions as in \eqref{micro_state}.

\subsection{Fokker–Planck description of the national trade}\label{QI_natio}
As it is well known in the study of kinetic models for socio-economic systems,\cite{albi2014boltzmann,bassetti2010explicit,bisi2022kinetic,pareschi2013interacting,pareschi2014wealth,toscani2006kinetic} deriving analytical properties of distribution functions directly from the Boltzmann equation \eqref{gen_kin_eq} is highly challenging. Nevertheless, for Boltzmann-type equations with collision-like structures, several analytical techniques have been developed to approximate long-time behavior and stationary distribution profiles $f^{\infty}$ through suitable asymptotic methods. Among these, one particularly effective approach is based on the framework of \emph{quasi-invariant interactions}.

This concept, first introduced in the kinetic theory of multi-agent systems\cite{cordier2005kinetic,toscani2006kinetic} and inspired by the notion of grazing collisions in classical kinetic theory,\cite{villani1998new} provides a systematic way to approximate the original integro-differential Boltzmann equation by a Fokker–Planck type partial differential equation, which is far more amenable to analytical study. The key idea is to introduce a small parameter $\varepsilon>0$ that quantifies the interaction strength: when $\varepsilon\to0^+$, agents exchange only a negligible amount of wealth or knowledge in each interaction. In this quasi-invariant regime, the differences between post- and pre-interaction states remain small, and this enables the rigorous derivation of PDE approximations for the underlying kinetic system. In the following, we apply this quasi-invariant interaction approach to our Boltzmann model in order to derive Fokker–Planck type equations and study the asymptotic behavior of their solutions.

Let us assume that the random variables $\kappa$, $\mu$, and $\eta$ are independent, identically distributed, and possess bounded moments of order at least three. Considering the microscopic interaction rules \eqref{micro_know} and \eqref{micro_wealth} for knowledge and wealth, respectively, we obtain
\begin{align} 
&\langle x'-x \rangle=(\beta(v)-\lambda(x))x +\lambda_b(x)b=:A_x(\z ,b)\,,\\[0.3cm]
&\langle v'-v \rangle= \gamma(\psi(y)w-\psi(x)v) +\bar{\eta}\phi(x,v)=:A_v(\z ,\z_*)
\end{align}
and
\begin{align}
&\langle (x'-x)^2 \rangle=A_x^2(\z ,b)+\delta^2x^2\,,\\[0.3cm]
&\langle (v'-v)^2 \rangle=A_v^2(\z ,\z_*)+\sigma^2v^2+\phi^2(x,v)(\omega^2-\bar{\eta}^2)\,, \\[0.3cm]
&\langle (v'-v)(x'-x) \rangle=A_x(\z ,b)A_v(\z ,\z_*)\,.
\end{align}
Analogous expressions hold for the differences $\langle y'-y\rangle$ and $\langle w'-w\rangle$, obtained by exchanging $x$ with $y$, $v$ with $w$, and vice versa. Now, let $\varphi(\z')$ be a smooth test function, and expand it in a Taylor series around $\z$ up to second order. This gives
\begin{equation}\label{gen_exp_limit1}
    \begin{split}
&\langle \varphi(\z')-\varphi(\z)\rangle= A_x(\z,b)\dfrac{\partial}{\partial x}\varphi(\z )+A_v(\z ,\z_*)\dfrac{\partial}{\partial v}\varphi(\z )+\\[0.5cm]&\dfrac{1}{2}\left[\delta^2x^2\dfrac{\partial^2}{\partial x^2}\varphi(\z )\!+\!\left[\phi^2(x,v)(\omega^2-\bar{\eta}^2) + \sigma^2v^2\right]\dfrac{\partial^2}{\partial v^2}\varphi(\z )\right]+\\[0.3cm]
&\dfrac{1}{2}\left[A_x^2(\z,b)\dfrac{\partial^2}{\partial x^2}\varphi(\z )\!+\!A_v^2(\z ,\z_*)\dfrac{\partial^2}{\partial v^2}\varphi(\z )\!+\!A_x(\z,b)A_v(\z ,\z_*)\dfrac{\partial^2}{\partial x\partial v}\varphi(\z )\right]\!\!+\mathcal{R}(\z ,\z_*),
    \end{split}
\end{equation}
where $\mathcal{R}(\z ,\z_*)$ denotes the remainder of the Taylor expansion.

To derive a meaningful asymptotic limit, we introduce a small parameter ${\varepsilon>0}$ and prescribe the following scaling for the coefficients in the microscopic rules \eqref{micro_know}–\eqref{micro_wealth}:
\begin{equation}\label{scale_1}
\lambda(x)\to\varepsilon\lambda(x)\,,\qquad\lambda_b(x)\to\varepsilon\lambda_b(x)\,,\qquad\beta(v)\to\varepsilon\beta(v)\,,\qquad\gamma\to\varepsilon\gamma\,.
\end{equation}
For the centered random variables we impose
\begin{equation}\label{scale_2}
\kappa\to\sqrt{\varepsilon}\kappa\,,\qquad\mu\to\sqrt{\varepsilon}\mu\,,
\end{equation}
so that their variance turns out to be $O(\varepsilon)$,
while for the non-centered random variable we set
\begin{equation}\label{scale_3}
\eta\to\varepsilon\eta\,,
\end{equation}
so that its mean is $O(\varepsilon)$.
Considering the weak form of the kinetic equation \eqref{gen_kin_eq_sym}, we analyze the asymptotic limit $\varepsilon\to0^+$ via the second-order Taylor expansion \eqref{gen_exp_limit1} of the test function $\varphi(\z')$ around $\z$. Assuming $\chi=1$ and applying the scalings \eqref{scale_1}–\eqref{scale_2}–\eqref{scale_3}, we obtain
\begin{equation}\label{gen_exp_limit2}
    \begin{split}
&\langle \varphi(\z ')-\varphi(\z )\rangle=\\[0.3cm]
&\varepsilon\left[A_x(\z ,b)\dfrac{\partial}{\partial x}\varphi(\z )+A_v(\z ,\z_*)\dfrac{\partial}{\partial v}\varphi(\z )+\dfrac{1}{2}\left[\delta^2x^2\dfrac{\partial^2}{\partial x^2}\varphi(\z )+\sigma^2v^2 \dfrac{\partial^2}{\partial v^2}\varphi(\z )\right]\right]+\\[0.3cm]
&\dfrac{\varepsilon^2}{2}\left[A_x^2(\z ,b)\dfrac{\partial^2}{\partial x^2}\varphi(\z )+\left[A_v^2(\z,\z_*)+\phi^2(x,v)(\omega^2-\bar{\eta}^2)\right]\dfrac{\partial^2}{\partial v^2}\varphi(\z )\right]+\\[0.3cm]
&\dfrac{\varepsilon^2}{2}\left[A_x(\z ,b)A_v(\z,\z_*)\dfrac{\partial^2}{\partial x\partial v}\varphi(\z)\right]+\mathcal{R}_\varepsilon(\z ,\z_*)\,,
    \end{split}
\end{equation}
where $\mathcal{R}_\varepsilon(\z ,\z)$ denotes the rescaled remainder, depending multiplicatively on higher moments of $\sqrt{\varepsilon}\kappa$, $\sqrt{\varepsilon}\mu$, and $\varepsilon\eta$. Hence $\mathcal{R}_\varepsilon(\z ,\z*)\ll1$ for $\varepsilon\ll1$ (a rigorous detailed proof that the remainder vanishes in proper quasi--invariant limits may be found in Ref.~\cite{cordier2005kinetic}). By setting $\tau=\varepsilon t$ and $f(t,\z)=g_\varepsilon(\tau,\z)$, we obtain that $g_\varepsilon$ satisfies
\begin{equation}
\begin{split}
    &\dfrac{d}{d\tau}\int\limits_{\mathbb{R}^2_+}\varphi(\z )g_\varepsilon(\tau,\z )d\z =\\[0.2cm]
    &\int\limits_{\mathbb{R}^2_+}\left[\mathcal{A}_x(\z )\dfrac{\partial \varphi}{\partial x}+\mathcal{A}_v(\tau,\z )\dfrac{\partial\varphi}{\partial v}+\dfrac{1}{2}\delta^2x^2\dfrac{\partial^2\varphi}{\partial x^2}+\dfrac{1}{2}\sigma^2v^2\dfrac{\partial^2\varphi}{\partial v^2}\right]g_\varepsilon(\tau,\z )d\z +\tilde{\mathcal{R}}_\varepsilon(\varphi)\,,
\end{split}
\end{equation}
with
\begin{align}
    &\mathcal{A}_x(\z)=\int\limits_{\mathbb{R}_+}A_x(\z ,b)C(b)db=(\beta(v)-\lambda(x))x +\lambda_b(x)\bar{B} \label{drift_Ax}\\[0.3cm] 
    &\mathcal{A}_v(\tau,\z)=\int\limits_{\mathbb{R}_+^2}A_v(\z,\z_*)g_\varepsilon(\tau,\z_*)d\z_*=\bar{\eta}\phi(x,v)-\gamma\psi(x)v +\gamma \int\limits_{\mathbb{R}_+^2}w\psi(y)g_\varepsilon(\tau,\z_*)d\z_* \label{drift_Av}
\end{align}
and a remainder
\begin{equation}
    \begin{split}
\tilde{\mathcal{R}}_\varepsilon(\varphi)=&\dfrac{\varepsilon}{2}\!\int\limits_{\mathbb{R}_+^5}\!\left[A_x^2(\z ,b)\dfrac{\partial^2\varphi}{\partial x^2}\!+\!A_v^2(\z ,\z_*)\dfrac{\partial^2\varphi}{\partial v^2}\right]g_\varepsilon(\tau,\z )g_\varepsilon(\tau,\z_*)C(b)dbd\z d\z_*+\\[0.3cm]
&\dfrac{\varepsilon}{2}\!\int\limits_{\mathbb{R}_+^5}\phi^2(x,v)(\omega^2\!-\!\bar{\eta}^2)\dfrac{\partial^2\varphi}{\partial v^2}g_\varepsilon(\tau,\z )g_\varepsilon(\tau,\z_*)C(b)dbd\z d\z_*+\\
&\dfrac{\varepsilon}{2}\int\limits_{\mathbb{R}_+^5}A_x(\z ,b)A_v(\z ,\z_*)\dfrac{\partial^2\varphi}{\partial x\partial v}g_\varepsilon(\tau,\z_*)C(b)dbd\z d\z_*+\\[0.3cm]
&\dfrac{1}{\varepsilon}\int\limits_{\mathbb{R}_+^5}\mathcal{R}_\varepsilon(\z ,\z_*)g_\varepsilon(\tau,\z )g_\varepsilon(\tau,\z_*)C(b)dbd\z d\z_*
    \end{split}
\end{equation}
that vanishes as $\varepsilon\to0^+$ (see Ref. \cite{cordier2005kinetic}). Passing to the limit $\varepsilon\to0^+$, the density $g_\varepsilon(\tau,\z )$ converges to $g(\tau,\z)$, which satisfies the equation
\begin{equation}
    \dfrac{d}{d\tau}\int\limits_{\mathbb{R}^2_+}\varphi(\z )g(\tau,\z)d\z =\!\!\!\int\limits_{\mathbb{R}^2_+}\left[\mathcal{A}_x(\z)\dfrac{\partial \varphi}{\partial x}\!+\!\mathcal{A}_v(\tau,\z)\dfrac{\partial\varphi}{\partial v}\!+\!\dfrac{1}{2}\delta^2x^2\dfrac{\partial^2\varphi}{\partial x^2}\!+\!\dfrac{1}{2}\sigma^2v^2\dfrac{\partial^2\varphi}{\partial v^2}\right]g(\tau,\z)d\z 
\end{equation}
that is the weak form of the Fokker–Planck equation
\begin{equation}\label{FP}
\begin{split}
   \dfrac{\partial}{\partial\tau}g(\tau,\z)=&-\dfrac{\partial}{\partial x}(\mathcal{A}_x(\z)g(\tau,\z))-\dfrac{\partial}{\partial v}(\mathcal{A}_v(\tau,\z )g(\tau,\z))+\\[0.3cm]
   &\dfrac{1}{2}\left[\delta^2\dfrac{\partial^2}{\partial x^2}(x^2g(\tau,\z))+\sigma^2\dfrac{\partial^2}{\partial v^2}(v^2g(\tau,\z))\right]\,. 
   \end{split}
\end{equation}
The rigorous derivation of \eqref{FP} follows the arguments of Refs.~\cite{cordier2005kinetic,toscani2006kinetic,pareschi2014wealth}. We remark that assuming the functions $\lambda(x)$ and $\beta(v)$ are of the same order of magnitude as $\delta^2$ ensures that, in the limiting equation, the interplay between knowledge and wealth, mediated by the intensity functions $\lambda(x)$ and $\beta(v)$, is preserved, while the influence of randomness is retained through the variance of $\kappa$. Analogously, the balance $\sigma^2/\gamma=\sigma^2/\bar{\eta}=cost$ ensures that the limiting equation preserves the effects of saving, represented by $\gamma$, and of risk, captured through both the mean of $\eta$ and the variance of $\mu$. 
Finally, substituting \eqref{drift_Ax}–\eqref{drift_Av} into \eqref{FP} yields the explicit Fokker–Planck equation. 
\begin{equation}\label{FP_2}
\begin{split}
   \dfrac{\partial g(\tau,x,v)}{\partial\tau}=&\dfrac{\partial}{\partial x}\left[\left((\lambda(x)-\beta(v))x -\lambda_b(x)\bar{B} \right)g(\tau,x,v)\right]+\\[0.3cm]
   &\dfrac{\partial}{\partial v}\left[\left(\gamma\left(\psi(x)v -M_w^\psi(\tau)\right)-\bar{\eta}\phi(x,v)\right) g(\tau,x,v)\right]+\\[0.3cm]
   &\dfrac{1}{2}\left[\delta^2\dfrac{\partial^2(x^2g(\tau,x,v))}{\partial x^2}+\sigma^2\dfrac{\partial^2 (v^2g(\tau,x,v))}{\partial v^2}\right]\,, 
   \end{split}
\end{equation}
where
$$
M_w^\psi(\tau):=\left\langle\, \,\,\int\limits_{\mathbb{R}^2_+}w\psi(y)g(\tau,\z_*)d\z_*\,\,\right\rangle\,.
$$
If the mean of the random variable $\eta$ vanishes, i.e. $\bar{\eta}=0$, and also $\beta(v) = 0$\,\, $\forall v \in \mathbb{R}_+$, one recovers the kinetic model proposed in Ref.~\cite{pareschi2013interacting}. Nevertheless, despite its seemingly simpler structure compared to \eqref{gen_kin_eq_sym}, the analytic characterization of the steady state of \eqref{FP_2} remains challenging. In the following, we focus on a partial analysis of \eqref{FP_2}, considering separately the steady-state distributions with respect to the two microscopic variables, wealth $v$ and knowledge $x$.

Let denote the steady-state distributions with respect to the two microscopic variables $x$ and $v$ by $g^{\infty}_x(\z)$ and $g^{\infty}_v(\z)$, respectively. We investigate their possible expressions by separately nullifying the derivatives with respect to $x$ and $v$ in \eqref{FP}. Starting from the derivatives w.r.t. $v$, we get
\begin{equation}
\begin{split}
    -\dfrac{\partial}{\partial v}(\mathcal{A}_v(\tau,\z )g_v^{\infty}(\z ))+\dfrac{1}{2}\sigma^2\dfrac{\partial^2}{\partial v^2}(v^2g_v^{\infty}(\z ))&=0 \\[0.3cm]
    \Longleftrightarrow\,\, \left[\bar{\eta}\phi(x,v)-(\gamma\psi(x)+\sigma^2)v +\gamma M_w^\psi(\tau)\right]g_v^{\infty}(\z )&=\dfrac{1}{2}\sigma^2v^2\dfrac{\partial}{\partial v}(g_v^{\infty}(\z ))\,.
    \end{split}
\end{equation}
 Assuming that $\sigma^2v^2\ne0$ and $\phi(x,v)$ is linearly increasing with respect to $v$, i.e., $\phi(x,v)=\tilde{\phi}(x)v$, and integrating over $\mathbb{R}_+$ 
 w.r.t. $v$ we get
 
\begin{equation}\label{partial_v}
\begin{split}
    \int\dfrac{\partial_v g_v^{\infty}(\z )}{g_v^{\infty}(\z )}dv&=\int\dfrac{2\gamma M_w^\psi(\tau)}{\sigma^2}\dfrac{1}{v^2}dv-\int\dfrac{2}{v}\left[\dfrac{\gamma\psi(x)-\bar{\eta}\tilde{\phi}(x)}{\sigma^2}+1\right]dv\\[0.4cm]
     \Longleftrightarrow\,\,g_v^{\infty}(\z )&=\mathcal{C}_v\, v^{-2\left[\dfrac{\gamma\psi(x)-\bar{\eta}\tilde{\phi}(x)}{\sigma^2}+1\right]} \,\,\,\,e^{-\dfrac{2\gamma M_w^\psi(\tau)}{v\sigma^2}}
    \end{split}
\end{equation}
where $\mathcal{C}_v$ is a normalization constant. Since $\bar{\eta}$ is set to be sufficiently small, we can assume that the exponent $(\gamma\psi(x)-\bar{\eta}\phi(x))/\sigma^2+1$ is always positive. In this case, the distribution $g_v^{\infty}(\z )\to0$ when $v\to0^+$, while for $v\to +\infty$ the function $g_v^{\infty}(\z )$ is asymptotic to an inverse power law $v^{-(1+I_v)}$  with $I_v$ representing the Pareto index,\cite{gualandi2018pareto} given by 
\begin{equation}
I_v=\dfrac{2(\gamma\psi(x)-\bar{\eta}\tilde{\phi}(x))}{\sigma^2}+1\,.
\end{equation}
Our model is thus able to reproduce the fact that wealth distributions usually have ``fat tails'' (with a finite number of convergent moments), meaning that there is a small fraction of population keeping the major part of total wealth.\cite{cordier2005kinetic,gualandi2018pareto} We note that the assumption of a proper perception of risk increasing with respect to individual wealth ($\phi(x,v)=\tilde{\phi}(x)v$), reasonable since rich people have more opportunities to improve their market awareness, provides a decrease in the Pareto index, consistently with the fact that rich people are favoured by knowledge.

Analogously, by considering the derivatives with respect to $x$, and assuming the simple case in which the personal selection rate $\lambda(x)$ and the external knowledge acquisition rate $\lambda_b(x)$ are constant, integration with respect to $x$ yields
\begin{equation}\label{partial_x}
  g_x^{\infty}(\z )= \mathcal{C}_x\,x^{-2\left[\dfrac{\lambda-\beta(v)}{\delta^2}+1\right]} \,\,\,\,e^{-\dfrac{2\lambda_b\bar{B}}{x\delta^2}}\,,
\end{equation}
with $\mathcal{C}_x$ normalization constant. Since we assume $\lambda_{-} > \beta_{+}$ in order to preserve the boundedness of the mean knowledge, the exponent $(\lambda-\beta(v))/\delta^2+1$ is always positive. Consequently, the distribution $g_x^{\infty}(\z)$ vanishes as $x \to 0^+$, while for ${x \to +\infty}$ it behaves asymptotically like an inverse power law $x^{-(1+I_x)}$. In particular, we can define a new index, which is analogous to the classical Pareto index, given by  \begin{equation}I_x=\dfrac{2(\lambda-\beta(v))}{\delta^2}+1\,.\end{equation} This result highlights that the model is consistent with the emergence, within society, of a very small class of highly learned individuals.\cite{gualandi2018pareto}

\begin{oss*}
In general, $g(\tau,\z)$ depends on both $x$ and $v$ and cannot be factorized. In such cases, only conditional distributions such as $g(\tau,\z)=g(\tau,x)g(\tau,v|x)$ or $g(\tau,\z)=g(\tau,v)g(\tau,x|v)$ are meaningful. However, if we assume
\begin{equation}\label{factor_g}
g(\tau,\z)=g_x(\tau,x)\,g_v(\tau,v),
\end{equation}
substitution into the Fokker--Planck equation \eqref{FP} leads, at steady state, to
\begin{equation}\label{factorized_ST}
-\frac{\partial_x\big(g_x^\infty(x)\mathcal{A}_x(\z)\big)}{g_x^\infty(x)}+\frac{\delta^2}{2}\frac{\partial_{xx}^2(x^2 g_x^\infty(x))}{g_x^\infty(x)}=\frac{\partial_v\big(g_v^\infty(v)\mathcal{A}_v(\tau,\z)\big)}{g_v^\infty(v)}-\frac{\sigma^2}{2}\frac{\partial_{vv}^2(v^2 g_v^\infty(v))}{g_v^\infty(v)}.
\end{equation}

If $\mathcal{A}_x=\mathcal{A}_x(x)$ (or $\mathcal{A}_v=\mathcal{A}_v(\tau,v)$), the two sides of \eqref{factorized_ST} depend on different variables and must therefore equal a constant $K$. Standard integrability arguments show that $K=0$, otherwise the solution diverges. Hence,
\begin{equation}
\begin{split}
&-\frac{\partial_x(g_x^\infty(x)\mathcal{A}_x(x))}{g_x^\infty(x)}+\frac{\delta^2}{2}\frac{\partial_{xx}^2(x^2 g_x^\infty(x))}{g_x^\infty(x)}=0,\\[0.2cm]
&-\frac{\partial_v(g_v^\infty(v)\mathcal{A}_v(\tau,v))}{g_v^\infty(v)}+\frac{\sigma^2}{2}\frac{\partial_{vv}^2(v^2 g_v^\infty(v))}{g_v^\infty(v)}=0,
\end{split}
\end{equation}
whose solutions coincide with the partial steady-state equilibria found above. Specifically, this holds provided $\beta(v)$ is constant (in the knowledge-related equation) or $\phi(x),\psi(x)$ are constant (in the wealth-related equation). Therefore partial equilibria can be seen as the natural consequence of the factorization \eqref{factor_g}, linking them to the independence of $x$ and $v$ in the limit.
\end{oss*}

\section{The Boltzmann model for international markets with individual transfer}
In several econophysics problems, it is natural to investigate the dynamics of multiple interacting populations distributed across different countries. 
The evolution of wealth distribution in each country is influenced both by domestic trade within the country itself, and by international trade with other countries. In this section, we extend the framework introduced in Section \ref{national_trade} (with wealth interaction rule influenced by proper knowledge and vice versa) to a multi--population system, incorporating also the possibility of agent transfers from one country to another.

In this setting, each agent is described by two continuous variables, namely wealth $v$ and knowledge $x$, as well as by a discrete label identifying its subgroup (country). More precisely, here an agent is represented by ${\z=(x,v,i)\in\mathbb{R}^2_+\times\mathcal{I}_n}$ with ${\mathcal{I}_n=\{1,\dots,n\}}$, where the index $i$ specifies the subgroup to which the agent belongs. A change in this label corresponds to a migration event (or transfer) between subgroups. Binary interactions between agents may trigger the transfers, but in such a way that the total mass in the system (i.e., the total number of individuals) is conserved. To formalize this mechanism, we introduce the conditional probability
\begin{equation}\label{prob_label}
P_{ik}^{i'k'}:=P\left((i',k')|(i,k)\right)  
\end{equation}
which denotes the probability that, after an interaction, a pair of agents initially in groups $(i,k)$ is transferred to groups $(i',k')$. Since the variables $i,k$ are discrete, the mapping $(i',k')\to P\left((i',k')|(i,k)\right)$ defines a discrete probability measure, and thus
\begin{equation}
\sum\limits_{i',k'=1}^nP\left((i',k')|(i,k)\right)=1\,.
\end{equation}
Given two interacting agents with microscopic states $\z=(x,v,i)$ and $\z_*=(y,w,k)$, we extend the interaction rules \eqref{micro_know} and \eqref{micro_wealth} to the case of multiple countries. The knowledge update is given by
\begin{equation}\label{micro_know_inter}
    x'=(1-\lambda_i(x))x+\lambda_b^i(x)b+\beta_i(v)x+\kappa_ix
\end{equation}
where the functions $\lambda_i(x)$, $\lambda_b^i(x)$, and $\beta_i(v)$ may differ across subgroups, reflecting population-specific learning and adaptation mechanisms. The stochastic perturbations $\kappa_i$ satisfy $\langle\kappa_i\rangle=0$ and $\langle\kappa_i^2\rangle=\delta_i^2$, for $i=1,..,n$. Similarly, the wealth update is defined as
\begin{equation}\label{micro_wealth_inter}
  \begin{sistem}
        v'=\left(1-\gamma_i\psi_i(x)\right)v+\gamma_k\psi_k(y)w+\phi_i(x,v)\eta_{ik}+\mu_{ik}v\\[0.3cm]
        w'=\left(1-\gamma_k\psi_k(y)\right)w+\gamma_i\psi_i(x)v+\phi_k(y,w)\eta_{ki}+\mu_{ki}w\,,
    \end{sistem}  
\end{equation}
where the exchange structure parallels the national trade model, but the functions $\psi_i(x)$ and $\phi_i(x)$ may vary across countries. The random variables $\eta_{ik}$ and $\mu_{ik}$ depend on the interacting populations; indeed, unexpected fluctuations in the market could be strongly influenced by the main features of interacting countries (for instance, trade between a rich and a poor country could be very risky for the poor one). We assume that $\langle\mu_{ik}\rangle=0$, $\langle\mu^2_{ik}\rangle=\sigma^2_{ik}$ and $\langle\eta_{ki}\rangle=\bar{\eta}$, $\langle\eta^2_{ki}\rangle=\omega^2_{ik}$, for $i,k=1,..,n$.
Notice that we have modeled the interactions at a microscopic level only, while different approaches for a description of behavioral economy for a multi-population system take into account also {\it micro-macro} interactions, namely the effects on single individuals due to each country as a whole.\cite{bellomo2020swarms,Ko2023collective} In our model the action of the countries may be considered included in the external source of knowledge (background) and in random variables.

Assuming that the microscopic rule \eqref{micro_know_inter} governs the evolution of knowledge for an agent $\z$ belonging to the $i$-th population, and that the binary international trade rule \eqref{micro_wealth_inter} describes the wealth exchange between agents $\z$ and $\z_*$ from populations $i$ and $k$, respectively, the joint dynamics of wealth and knowledge for agents in population $i$ can be represented by the density function $f=f(t,\z)$. Precisely, $f(t,x,v,i)dxdv$ denotes the proportion of agents with label $i\in\mathcal{I}_n$ whose microscopic state lies between $x$ and $x+dx$ and $v$ and $v+dv$ at time $t$. Since $i$ is a discrete variable, we may represent $f$ as in Ref.~\cite{bisi2024kinetic}, namely
\begin{equation}
f(t,\z)=f(t,x,v,i)=\sum\limits_{k=1}^nf_i(t,x,v)\delta(i-k)
\end{equation}
where $f_i=f_i(t,x,v)\ge0$ is the distribution function of the microscopic state $(x,v)$ for the agents with label $i$. 
Because both the interactions and the label-switching dynamics preserve the total mass of the system, we may assume that $f=f(t,\z)$ is a probability distribution, i.e.,
\begin{equation}
\int\limits_{\mathbb{R}^2_+}\int\limits_{\mathcal{I}_n}f(t,x,v,i)dxdvdi=\sum\limits_{i=1}^n\int\limits_{\mathbb{R}^2_+}f_i(t,x,v)dxdv=1\qquad \forall t>0\,.
\end{equation}
The distributions $f_i(t,x,v)$, for $i=1,..,n$ are in general not probability density functions since their integral over the microscopic state $(x,v)$ is not preserved in time due to label switching. We 
denote by 
\begin{equation}
\rho_i(t):=\int\limits_{\mathbb{R}^2_+}f_i(t,x,v)dxdv
\end{equation}
the total mass of the subgroup of agents with label $i$ at time $t$. Clearly, $0\le \rho_i(t)\le1$ and 
\begin{equation}
\sum\limits_{i=1}^n\rho_i(t)=1\qquad\forall t>0\,.
\end{equation}
Furthermore, we define the first moment of $f_i$ with respect to the knowledge and wealth variables as
\begin{equation}
M_x^i(t):=\int\limits_{\mathbb{R}^2_+}xf_i(t,x,v)dxdv\,,\qquad M_v^i(t):=\int\limits_{\mathbb{R}^2_+}vf_i(t,x,v)dxdv\,.
\end{equation}
Hence, the average knowledge and wealth of the $i$th subgroup are given by
\begin{equation}
m_x^i(t):=\dfrac{M_x^i(t)}{\rho_i(t)}\,,\qquad m_v^i(t):=\dfrac{M_v^i(t)}{\rho_i(t)}\,.
\end{equation}

Let $\chi_{ik}$ denote the interaction frequency between an agent $\z$ in  population $i$ and an agent $\z_*$ in population $k$. We then define $\theta_{ik}^{i'k'}:=\chi_{ik}P_{ik}^{i'k'}$, where $P_{ik}^{i'k'}$ is the probability of label switching from $(i,k)$ to $(i',k')$, introduced in \eqref{prob_label}. Let then consider binary and symmetric interactions between the agents $\z=(x,v,i)$ and $\z_*=(y,w,k)$. Given a generic test function $\varsigma=\varsigma(x,v,i):\mathbb{R}^2_+\times\mathcal{I}_n\to\mathbb{R}$ and assuming a factorization $\varsigma(x,v,i)=\vartheta(i)\varphi(x,v)$  such that $\vartheta(s)=1$ for a certain $s\in\mathcal{I}_n$ and $\vartheta(i)=0$ for all $i\in\mathcal{I}_n\backslash\{s\}$
the time evolution of each density distribution $f_s(t,x,v)$, for $s=1,..,n$, is governed by a Boltzmann-type equation, which in weak form reads
\begin{equation}\label{gen_kin_eq_inter}\begin{split}
    &\dfrac{d}{dt}\int\limits_{\mathbb{R}^2_+}\varphi(x,v)f_s(t,x,v)dxdv =\\[0.2cm]
&\!\!\!\left\langle\sum\limits_{i,k,k'=1}^n\int\limits_{\mathbb{R}^5_+}\left(\theta_{ik}^{sk'}\varphi(x',v')f_{i}(t,x',v')\!-\!\theta_{sk}^{ik'}\varphi(x,v)f_s(t,x,v)\right) C(b)f_k(t,y,w)dxdv\,dydw \,db\right\rangle\!\!
    \end{split}
\end{equation}
Equation \eqref{gen_kin_eq_inter} expresses the weak formulation of the kinetic dynamics governing the subgroup distribution $f_s(t,x,v)$. The right-hand side represents the balance between gain and loss terms due to interactions with agents from all populations. More precisely, the first contribution accounts for agents entering subgroup $s$ after an interaction, while the second describes agents leaving subgroup $s$. The kernel $\theta_{ik}^{sk'}$ encodes both the interaction frequency $\chi_{ik}$ and the probability of a label switch $P_{ik}^{i'k'}$ in the particular case $i'=s$, since we are considering the individuals going to group~$s$.

Following the approach described in Section \ref{QI_natio}, we apply the quasi-invariant interaction method to the Boltzmann model \eqref{gen_kin_eq_inter} in order to derive a system of Fokker–Planck type equations characterizing international trade dynamics. Assume that the random variables $\kappa_{i}$, $\mu_{ik}$, and $\eta_{ik}$ are independent, identically distributed, and possess bounded moments of order at least three for all $i,k=1,..,n$. Considering the microscopic interaction rules \eqref{micro_know_inter} and \eqref{micro_wealth_inter} for knowledge and wealth exchange with international transfers, we obtain
\begin{align} 
&\langle x'-x \rangle=(\beta_i(v)-\lambda_i(x))x +\lambda^i_b(x)b=:A^i_x(x,v,b)\,,\\[0.3cm]
&\langle v'-v \rangle= \gamma_k\psi_k(y)w-\gamma_i\psi_i(x)v +\bar{\eta}\phi_i(x,v)=:A^{ik}_v(x,v,y,w)
\end{align}
and
\begin{align}
&\langle (x'-x)^2 \rangle=(A^i_x)^2(x,v,b)+\delta_i^2x^2\,,\\[0.3cm]
&\langle (v'-v)^2 \rangle=(A^{ik}_v)^2(x,v,y,w)+\sigma_{ik}^2v^2+\phi_i^2(x,v)(\omega_{ik}^2-\bar{\eta}^2)\,, \\[0.3cm]
&\langle (v'-v)(x'-x) \rangle=A^i_x(x,v,b)A^{ik}_v(x,v,y,w)\,.
\end{align}
Analogous expressions hold for the differences $\langle y'-y\rangle$ and $\langle w'-w\rangle$, obtained by exchanging $x$ with $y$, $v$ with $w$, and vice versa. 

Introducing the small parameter $\varepsilon>0$, we prescribe scalings for the coefficients and random variables in \eqref{micro_know_inter} and \eqref{micro_wealth_inter} analogous to \eqref{scale_1}–\eqref{scale_2}–\eqref{scale_3}, for ${i,k=1,..,n}$. Given a smooth test function $\varphi(x,v)$, its expansion in a Taylor series around $(x,v)$ up to second order has the same structure as in \eqref{gen_exp_limit1}. Fixing a certain $s\in\mathcal{I}_n$ and setting $\tau=\varepsilon t$ and $f_s(t,x,v)=g_s(\tau,x,v)$, we obtain that $g_s$ satisfies
\begin{equation}\label{QI_equation}
\begin{split}
&\dfrac{d}{d\tau}\int\limits_{\mathbb{R}^2_+}\varphi g_s(\tau)dxdv= \sum\limits_{i,k,k'=1}^n\,\int\limits_{\mathbb{R}^2_+}\theta_{ik}^{sk'}\left(\rho_k(\tau)\mathcal{A}^i_x\dfrac{\partial\varphi}{\partial x}+\mathcal{A}^{ik}_v(\tau)\dfrac{\partial\varphi}{\partial v}\right)g_i(\tau)dxdv\,+\\[0.2cm]
&\sum\limits_{i,k,k'=1}^n\,\int\limits_{\mathbb{R}^2_+}\theta_{ik}^{sk'}\left(\dfrac{1}{2}\rho_k(\tau)\dfrac{\partial^2\varphi}{\partial x^2}\delta_i^2x^2+\dfrac{1}{2}\rho_k(\tau)\dfrac{\partial^2\varphi}{\partial v^2}\sigma_{ik}^2v^2\right)g_i(\tau)dxdv\,+\\[0.2cm]
&\dfrac{1}{\varepsilon}\sum\limits_{i,k,k'=1}^n\,\rho_k(\tau)\int\limits_{\mathbb{R}^2_+}\varphi\left(\theta_{ik}^{sk'}g_i(\tau)-\theta_{sk}^{ik'}g_s(\tau)\right)dxdv+\tilde{\mathcal{R}}^s_\varepsilon(\varphi)\,,
\end{split}
\end{equation}
where, for readability, the dependence on microscopic variables has been neglected. In particular,
\begin{align}
    &\mathcal{A}^i_x=\mathcal{A}^i_x(x,v)=\int\limits_{\mathbb{R}_+}A^i_xC(b)db=(\beta_i(v)-\lambda_i(x))x +\lambda^i_b(x)\bar{B} \label{drift_Ax_i}\\[0.3cm] 
    &\mathcal{A}^{ik}_v(\tau)=\mathcal{A}^{ik}_v(\tau,x,v)=\int\limits_{\mathbb{R}_+^2}A^{ik}_vg_k(\tau)dydw=\\[-0.2cm] 
    &\qquad\qquad\left(\bar{\eta}\phi_i(x,v)-\gamma_i\psi_i(x)v\right)\rho_k(\tau) +\gamma_k \int\limits_{\mathbb{R}_+^2}w\psi_k(y)g_k(\tau)dydw \label{drift_Av_ik}
\end{align}
and the remainder term is given by
\begin{equation}
    \begin{split}
&\tilde{\mathcal{R}}^s_\varepsilon(\varphi)=\sum\limits_{i,k,k'=1}^n\dfrac{\varepsilon}{2}\int\limits_{\mathbb{R}_+^5}\theta_{ik}^{sk'}(A_x^i)^2\dfrac{\partial^2\varphi}{\partial x^2}C(b)g_i(\tau)g_k(\tau)dxdv\,dydw\,db\,+\\
&\sum\limits_{i,k,k'=1}^n\dfrac{\varepsilon}{2}\int\limits_{\mathbb{R}_+^5}\theta_{ik}^{sk'}\left[(A_v^{ik})^2+\phi_i^2(x,v)(\omega^2-\bar{\eta}^2)\right]\dfrac{\partial^2\varphi}{\partial v^2}C(b)g_i(\tau)g_k(\tau)dxdv\,dydw\,db\,+\\[0.3cm]
&\sum\limits_{i,k,k'=1}^n\dfrac{\varepsilon}{2}\int\limits_{\mathbb{R}_+^5}\theta_{ik}^{sk'}\left[A^i_xA^{ik}_v\dfrac{\partial^2\varphi}{\partial x\partial v}\right]C(b)g_i(\tau)g_k(\tau)dxdv\,dydw\,db\, +\\[0.3cm]
&\dfrac{1}{\varepsilon}\int\limits_{\mathbb{R}_+^5}\theta_{ik}^{sk'}\mathcal{R}^{ik}_\varepsilon(x,v,y,w)C(b)g_i(\tau)g_k(\tau)dxdv\,dydw db
    \end{split}
\end{equation}
where $\mathcal{R}^{ik}_\varepsilon$ denotes the remainder of the Taylor expansion and it depends multiplicatively on higher moments of $\sqrt{\varepsilon}\kappa_i$, $\sqrt{\varepsilon}\mu_{ik}$, and $\varepsilon\eta_{ik}$, for $i,k=1,..,n$. 
We now consider an asymptotic expansion of the probability density function of the whole population $g$, i.e.,
\begin{equation}
g(\tau,x,v,i)=g^{(0)}(\tau,x,v,i)+\varepsilon g^{(1)}(\tau,x,v,i)+\mathcal{O}(\varepsilon^2)\,,
\end{equation}
where the zero-th term $g^{(0)}$ is assumed to preserve the total mass. 
Specifically, preservation of mass by the zero-th order term means
\begin{equation}
\rho^{(0)}(\tau):=\int\limits_{\mathcal{I}_n}\int\limits_{\mathbb{R}^2_+}g^{(0)}(\tau,x,v,i)dxdvdi=\int\limits_{\mathcal{I}_n}\int\limits_{\mathbb{R}^2_+}g(\tau,x,v,i)dxdvdi\,,
\end{equation}
while the first-order correction has vanishing mass
\begin{equation}\label{rho1_assum}
\quad \rho^{(1)}(\tau):=\int\limits_{\mathcal{I}_n}\int\limits_{\mathbb{R}^2_+}g^{(1)}(\tau,x,v,i)dxdvdi=0\,.
\end{equation}
In particular,
$$\rho^{(0)}=\sum_{i=1}^n\rho_i^{(0)}\,.$$ 
Therefore, each subgroup distribution $g_i$ can be expanded as
\begin{equation}\label{gi_exp}
g_i(\tau,x,v)=g_i^{(0)}(\tau,x,v)+\varepsilon g_i^{(1)}(\tau,x,v)+\mathcal{O}(\varepsilon^2)\,.
\end{equation}
Plugging \eqref{gi_exp} into \eqref{QI_equation}, we compare equal powers of $\varepsilon$, starting from the leading order terms:
\begin{equation}
    \sum\limits_{i,k,k'=1}^n\,\rho^{(0)}_k(\tau)\int\limits_{\mathbb{R}^2_+}\varphi \left(\theta_{ik}^{sk'}g_i^{(0)}(\tau,x,v)-\theta_{sk}^{ik'}g_s^{(0)}(\tau,x,v)\right)dxdv=0\,.
\end{equation}
The $\varepsilon^0$-term yields the fast label–switching equilibrium, which expresses the instantaneous redistribution of agents across countries:
\begin{equation}\label{g0_s_QI}
    g_s^{(0)}(\tau,x,v)=\dfrac{\sum\limits_{\substack{i,k,k'=1\\i\ne s}}^n\theta_{ik}^{sk'}g_i^{(0)}(\tau,x,v)\rho^{(0)}_k(\tau)}{\sum\limits_{\substack{i,k,k'=1\\i\ne s}}^n\theta_{sk}^{ik'}\rho^{(0)}_k(\tau)}.
\end{equation}
This relation should be interpreted as a compatibility condition: the leading-order densities $g_s^{(0)}$ are not independent, but are coupled pointwise in $(x,v)$ through the fast label–switching mechanism. In other words, the system equilibrates instantaneously in the discrete variable $s$, while the continuous variables remain unchanged at this scale.
At the next order we obtain
\begin{equation}\label{QI_equation_first}
\begin{split}
    &\dfrac{d}{d\tau}\int\limits_{\mathbb{R}^2_+}\varphi g_s^{(0)}(\tau)dxdv= \\[0.2cm]
 &\sum\limits_{i,k,k'=1}^n\,\,\int\limits_{\mathbb{R}^2_+}\theta_{ik}^{sk'}\left(\rho^{(0)}_k(\tau)\mathcal{A}^i_x\dfrac{\partial\varphi}{\partial x}+\mathcal{A}^{ik}_v(\tau)\dfrac{\partial\varphi}{\partial v}\right)g_i^{(0)}(\tau)dxdv+\\[0.2cm]  &\sum\limits_{i,k,k'=1}^n\,\,\int\limits_{\mathbb{R}^2_+}\theta_{ik}^{sk'}\left(\dfrac{1}{2}\rho^{(0)}_k(\tau)\dfrac{\partial^2\varphi}{\partial x^2}\delta_i^2x^2+\dfrac{1}{2}\rho^{(0)}_k(\tau)\dfrac{\partial^2\varphi}{\partial v^2}\sigma_{ik}^2v^2\right)g_i^{(0)}(\tau)dxdv+\\[0.2cm]
    &\sum\limits_{i,k,k'=1}^n\,\rho_k^{(0)}(\tau)\int\limits_{\mathbb{R}^2_+}\varphi\left(\theta_{ik}^{sk'}g_i^{(1)}(\tau)-\theta_{sk}^{ik'}g_s^{(1)}(\tau)\right)dxdv\,,
    \end{split}
\end{equation}
where we have used assumption \eqref{rho1_assum}. For $s\in\mathcal{I}_n$, integrating by parts and rewriting in strong form leads to the Fokker–Planck system of equations 
\begin{equation}\label{FP_interna}
\begin{split}
   & \dfrac{d}{d\tau} g_s^{(0)}(\tau,x,v)=\\[0.2cm]
   &\!-\dfrac{\partial}{\partial x}\! \left(\sum\limits_{i,k,k'=1}^n\!\! \theta_{ik}^{sk'}\mathcal{A}^i_x(x,v)g_i^{(0)}(\tau,x,v)\rho_k^{(0)}(\tau)\!\right)\!\!-\!\dfrac{\partial}{\partial v} \!\left(\sum\limits_{i,k,k'=1}^n \!\!\theta_{ik}^{sk'}\mathcal{A}^{ik}_v(\tau,x,v)g_i^{(0)}(\tau,x,v)\!\right)
    \\[0.2cm]
    &+\dfrac{1}{2}\sum\limits_{i,k,k'=1}^n\theta_{ik}^{sk'}\rho_k^{(0)}(\tau)\left[\dfrac{\partial^2}{\partial x^2} \left( \delta_i^2x^2g_i^{(0)}(\tau,x,v)
    \!\right)+\dfrac{\partial^2}{\partial v^2}\left(\sigma_{ik}^2v^2g_i^{(0)}(\tau,x,v)\right)\right]\\[0.2cm]
    &+\sum\limits_{i,k,k'=1}^n\,\rho_k^{(0)}(\tau)\left(\theta_{ik}^{sk'}g_i^{(1)}(\tau,x,v)-\theta_{sk}^{ik'}g_s^{(1)}(\tau,x,v)\right)
     \end{split}
\end{equation}
for $s=1,..,n$. The weak formulation \eqref{QI_equation_first} is equivalent, after integration by parts, to the coupled Fokker–Planck system \eqref{FP_interna}. This system governs the slow drift–diffusion in knowledge and wealth under the constraint of fast label–switching equilibrium. The resulting structure clearly illustrates the time-scale separation: the label dynamics equilibrate instantaneously at order $\varepsilon^0$, while the continuous dynamics emerge only at order $\varepsilon^1$. Such decompositions are classical in kinetic theory, and provide the effective macroscopic description of the process. Moreover, we observe that by fixing $n=1$, and hence $s=1$, the label-switch term disappears, and the system \eqref{FP_interna} reduces to the Fokker–Planck equation \eqref{FP}.

\subsection{Macroscopic equations of a two-country economy}\label{2C_economy}
Building on the kinetic model introduced in the previous section, we now specialize the formulation to a two-population setting ($n=2$), where binary interactions simultaneously govern the exchange of wealth between the two populations, the evolution of individual knowledge, and the migration of agents across countries through label switching. The objective of this section is to derive the evolution equations for the population sizes $\rho_i$, as well as for the mean wealth $m_v^i$ and mean knowledge $m_x^i$ of each $i$-population, for $i=1,2$.

To specify the formulation of the model, we distinguish between two types of interactions: (i) domestic trade, where agents belong to the same country, and (ii) international trade, where agents belong to different countries. Each interaction may occur either without migration or with migration of one of the agents involved. In particular, we allow for transfers where (a)-(b) an interaction between two individuals of the same country results in the migration of one of them to the other country, or (c)-(d) an interaction between agents from different countries leads both agents to decide to reside in the same country. The only restriction imposed is that at most one agent can migrate in each interaction. This assumption leads to the following admissible transfer mechanisms:
\begin{equation}
    \begin{split}
        \text{(a)}\,\,\,\, 1+1\to1+2\qquad\qquad\text{(b)}\,\,\,\, 2+2\to1+2\\[0.2cm]
        \text{(c)}\,\,\,\, 1+2\to1+1\qquad\qquad\text{(d)}\,\,\,\, 1+2\to2+2
    \end{split}
\end{equation}
Translating these assumptions into the interaction probabilities between agents, we find that the only non-vanishing values of $P_{ik}^{i'k'}$ correspond to the following cases.
\begin{itemize}
    \item {\bf Domestic and international trade without migration}:
    \begin{equation}\label{admis_comm}
    (i,k,i',k')\in\{(1,1,1,1),(2,2,2,2),(1,2,1,2),(2,1,2,1)\}
\end{equation}
which represent interactions that preserve both labels.
\item {\bf Interactions with migration of a single agent}:
\begin{equation}\label{admis_transf}
\begin{split}
    (i,k,i',k')\in\{&(1,1,1,2),(1,1,2,1),(2,2,1,2),(2,2,2,1),(1,2,1,1),\\[0.2cm]
    &(1,2,2,2),(2,1,1,1),(2,1,2,2)\}
    \end{split}
\end{equation}
that correspond to either domestic interactions leading to the transfer of one agent to the other country, or international interactions in which both agents end up residing in the same country.
\end{itemize}

Considering the microscopic dynamics \eqref{micro_know_inter}–\eqref{micro_wealth_inter} governing knowledge and wealth evolution, together with the interaction probabilities $\theta_{ik}^{i'k'}$ defined in \eqref{admis_comm}–\eqref{admis_transf}, we employ the kinetic equation \eqref{gen_kin_eq_inter} to derive the macroscopic evolutionary equations of population sizes, mean wealth, and mean knowledge in each subgroup. In particular, by choosing the test function $\varphi(x,v)=1$ in \eqref{gen_kin_eq_inter}, and focusing on the cases $s=1,2$ we obtain the balance equations for the masses $\rho_i(t)$ of the two populations, which take the form
\begin{equation}\label{macro_rho}
    \begin{sistem}
        \dfrac{d}{dt}\rho_1(t)=\left(\theta_{21}^{11}\rho_2(t)-\theta_{11}^{21}\rho_1(t)\right)\rho_1(t)+\left(\theta_{22}^{12}\rho_2(t)-\theta_{12}^{22}\rho_1(t)\right)\rho_2(t)\\[0.4cm]
        \dfrac{d}{dt}\rho_2(t)=\left(\theta_{11}^{21}\rho_1(t)-\theta_{21}^{11}\rho_2(t)\right)\rho_1(t)+\left(\theta_{12}^{22}\rho_1(t)-\theta_{22}^{12}\rho_2(t)\right)\rho_2(t)\,.
    \end{sistem}
\end{equation}
These equations describe how migration, driven by the interaction rates $\theta_{ik}^{i'k'}$ redistributes the relative sizes of the two populations. Clearly, interactions corresponding to domestic or international trade without agent transfer do not affect the population sizes. By construction, the total population is conserved:
\begin{equation}
\dfrac{d}{dt}(\rho_1(t)+\rho_2(t))=0\,\qquad\Longrightarrow \bar{\rho}=\rho_1(t)+\rho_2(t)=1.
\end{equation}
Seeking a stationary configuration for the population masses, we obtain the relation
\begin{equation}\label{alpha}
\rho_2^\infty=\alpha\rho_1^\infty\qquad\text{with }\,\,\alpha=\dfrac{-(\theta_{21}^{11}-\theta_{12}^{22})+\sqrt{\left(\theta_{21}^{11}-\theta_{12}^{22}\right)^2+4\theta_{11}^{21}\theta_{22}^{12}}}{2\theta_{22}^{12}}
\end{equation}
which expresses the asymptotic ratio between the two populations. Depending on the specific values of the transition rates $\theta_{ik}^{i'k'}$, this ratio $\alpha$ can be either greater or smaller than one, indicating a long-term dominance of one population over the other. Since the total population is fixed, i.e., $\rho_1(t)+\rho_2(t)=\bar{\rho}$, the unique equilibrium distribution of agents between the two subgroups is given by
\begin{equation}\label{rhoi_infty}
\rho_1^\infty=\dfrac{\bar{\rho}}{1+\alpha}\,\qquad \rho_2^\infty=\dfrac{\alpha\bar{\rho}}{1+\alpha}\,.
\end{equation}
These results are consistent with the findings in Ref.~\cite{bisi2022kinetic}. In particular, when the probability that interacting agents with different labels migrate to the same subgroup is identical for both populations, i.e. $P_{12}^{11}=P_{12}^{22}$, and the interaction rates within each subgroup are symmetric, one obtains $\alpha=1$, leading to the symmetric equilibrium $\rho_1^\infty=\rho_2^\infty=\bar{\rho}/2$. Conversely, if $\theta_{12}^{11}=\theta_{12}^{22}$, but $P_{12}^{11}\ne P_{12}^{22}$ then $\alpha>1$ whenever the probability that an agent from group 1 migrates to group 2 exceeds that of the reverse migration, while
$\alpha<1$ holds in the opposite case. As a result, the long-term equilibrium may exhibit a reversal of the initial relative population sizes, as also discussed in Ref.~\cite{bisi2024kinetic}.

Next, we focus on the evolution of the total wealth $M_v^i(t)$ in each country, obtained by setting $\varphi(x,v)=v$ in the weak formulation \eqref{gen_kin_eq_inter}. In the following, we assume that $\phi_i(x,v)$ is linearly increasing with respect to $v$, i.e., $\phi_i(x,v)=\tilde{\phi}_i(x)v$, as done in Section \ref{QI_natio}. The corresponding macroscopic balance laws read
\begin{equation}\label{macro_Mv}
    \begin{sistem}
    \begin{aligned}
        \dfrac{d}{dt}M_v^1(t)=&\,\,\bar{\eta}\rho_1(t)M_1^\phi\left(\theta_{11}^{11}+\theta_{11}^{12}\right)+\bar{\eta}\rho_2(t)M_1^\phi\left(\theta_{12}^{12}+\theta_{12}^{11}\right)+ \\[0.3cm]
        &\bar{\eta}M_2^\phi\left(\theta_{21}^{11}\rho_1(t)+\theta_{22}^{12}\rho_2(t)\right)+\theta_{12}^{12}\left(\gamma_2\rho_1(t)M_2^\psi-\gamma_1\rho_2(t)M_1^\psi\right)+\\[0.3cm]
        &M_v^2\left(\theta_{21}^{11}\rho_1(t)+\theta_{22}^{12}\rho_2(t)\right)-M_v^1\left(\theta_{11}^{21}\rho_1(t)+\theta_{12}^{22}\rho_2(t)\right)\\[0.5cm]
        \end{aligned}\\[1.5cm]
        \begin{aligned}
        \dfrac{d}{dt}M_v^2(t)=&\,\,\bar{\eta}\rho_2(t)M_2^\phi\left(\theta_{22}^{22}+\theta_{22}^{21}\right)+\bar{\eta}\rho_1(t)M_2^\phi\left(\theta_{21}^{21}+\theta_{21}^{22}\right)+\\[0.3cm]
&\bar{\eta}M_1^\phi\left(\theta_{11}^{21}\rho_1(t)+\theta_{12}^{22}\rho_2(t)\right)+ \theta_{21}^{21}\left(\gamma_1\rho_2(t)M_1^\psi-\gamma_2\rho_1(t)M_2^\psi\right)+\\[0.3cm]
        &M_v^1\left(\theta_{11}^{21}\rho_1(t)+\theta_{12}^{22}\rho_2(t)\right)-M_v^2\left(\theta_{21}^{11}\rho_1(t)+\theta_{22}^{12}\rho_2(t)\right)\\[0.5cm] 
        \end{aligned}
    \end{sistem}
\end{equation}
where we have assumed symmetry in the interaction rates, i.e.,
\begin{equation}
\theta_{ii}^{12}=\theta_{ii}^{21}\,\qquad\theta_{12}^{ii}=\theta_{21}^{ii}\,,\qquad\forall i=1,2\,.
\end{equation}
In particular, we denote by
\begin{equation}
M_i^\phi(t):=\int\limits_{\mathbb{R}^2_+}v\tilde{\phi}_i(x)f_i(t,x,v)dxdv\,,\qquad M_i^\psi(t):=\int\limits_{\mathbb{R}^2_+}v\psi_i(x)f_i(t,x,v)dxdv\,.
\end{equation}
Note that, since in the microscopic rule \eqref{micro_wealth_inter} the random effects associated with risk-related fluctuations are assumed to have a small but positive mean, the total wealth $M_v^1+M^2_v$ is not conserved. Instead, it exhibits exponential growth in time. This reproduces the long-term historical trend where global wealth accumulates, though redistributed unevenly across countries.

By combining \eqref{macro_rho} and \eqref{macro_Mv}, we obtain the evolution equations for the mean wealths $m_v^i:=M_v^i/\rho_i$ of each country:
{\allowdisplaybreaks
\begin{equation}\label{macro_mv}
    \begin{sistem}
    \begin{aligned}
        \dfrac{d}{dt}m_v^1(t)=&\bar{\eta}\rho_1(t)m_1^\phi\left(\theta_{11}^{11}+\theta_{11}^{12}\right)+\bar{\eta}\rho_2(t)m_1^\phi\left(\theta_{12}^{12}+\theta_{12}^{11}\right) +\\[0.3cm]
        &\bar{\eta}\dfrac{\rho_2(t)}{\rho_1(t)}m_2^\phi\left(\theta_{21}^{11}\rho_1(t)+\theta_{22}^{12}\rho_2(t)\right)+\dfrac{\rho_2(t)}{\rho_1(t)}\left(\theta_{22}^{12}\rho_2(t)+\theta_{21}^{11}\rho_1(t)\right)\left(m_v^2-m_v^1 \right)+\\[0.3cm]
        &\theta_{12}^{12}\left(\gamma_2m_2^\psi-\gamma_1m_1^\psi\right)\rho_2(t)\\[0.5cm]
        \end{aligned}\\[1.5cm]
       \begin{aligned}
        \dfrac{d}{dt}m_v^2(t)=&\bar{\eta}\rho_2(t)m_2^\phi\left(\theta_{22}^{22}+\theta_{22}^{21}\right)+\bar{\eta}\rho_1(t)m_2^\phi\left(\theta_{21}^{21}+\theta_{21}^{22}\right) +\\[0.3cm]
        &\bar{\eta}\dfrac{\rho_1(t)}{\rho_2(t)}m_1^\phi\left(\theta_{11}^{21}\rho_1(t)+\theta_{12}^{22}\rho_2(t)\right)+\dfrac{\rho_1(t)}{\rho_2(t)}\left(\theta_{11}^{21}\rho_1(t)+\theta_{12}^{22}\rho_2(t)\right)\left(m_v^1-m_v^2 \right)+\\[0.3cm]
        &\theta_{21}^{21}\left(\gamma_1m_1^\psi-\gamma_2m_2^\psi\right)\rho_1(t)\,.\\
        \end{aligned}
    \end{sistem}
\end{equation}
}
Here, $m_i^\phi(t):=M_i^\phi(t)/\rho_i(t)$ and $m_i^\psi(t):=M_i^\psi(t)/\rho_i(t)$. It is evident that system \eqref{macro_mv} is not closed, since its evolution equations involve additional moments of the distribution functions $f_i(t,x,v)$, for $i=1,2$, denoted by $m_i^\phi(t)$ and $m_i^\psi(t)$. These equations describe how the mean wealth of each population is influenced by three main factors: endogenous wealth growth, which is knowledge-dependent and captured by the moments $m_i^\phi$; wealth exchanges resulting from domestic and international trade; and redistribution effects induced by migration flows between the two countries. In particular, if one country is wealthier on average, migration and trade tend to reduce wealth disparities, as reflected in the terms proportional to $(m_v^i-m_v^k)$, weighted by the ratio of the population sizes. Overall, system \eqref{macro_mv} describes scenarios where countries may either converge to a balanced state with comparable mean wealth or diverge, depending on the relative strength of endogenous growth versus redistribution through trade and migration. 

Assuming for simplicity that $\psi(x)=\bar{\psi}$ and $\tilde{\phi}(x)=\bar{\phi}$, and further setting ${\bar{\psi}=\bar{\phi}=1}$, system \eqref{macro_mv} reduces to a closed set of equations for the mean wealths $m_v^1(t)$ and $m_v^2(t)$. Moreover, if we neglect the contribution of the risk-related fluctuation term $\bar{\eta}$, the system simplifies to
\begin{equation}\label{macro_mv_red}
    \begin{sistem}
    \begin{aligned}
        \dfrac{d}{dt}m_v^1(t)=
        \dfrac{\rho_2(t)}{\rho_1(t)}\left(\theta_{22}^{12}\rho_2(t)+\theta_{21}^{11}\rho_1(t)\right)\left(m_v^2-m_v^1 \right)+\theta_{12}^{12}\left(\gamma_2m_v^2-\gamma_1m_v^1\right)\rho_2(t)\\[0.5cm]
        \end{aligned}\\
       \begin{aligned}
        \dfrac{d}{dt}m_v^2(t)=\dfrac{\rho_1(t)}{\rho_2(t)}\left(\theta_{11}^{21}\rho_1(t)+\theta_{12}^{22}\rho_2(t)\right)\left(m_v^1-m_v^2 \right)+\theta_{21}^{21}\left(\gamma_1m_v^1-\gamma_2m_v^2\right)\rho_1(t)\,.\\
        \end{aligned}
    \end{sistem}
\end{equation}
This system reproduces, as a particular case, the results in Ref.~\cite{bisi2022kinetic} under the simplifying assumptions $\theta_{22}^{12}=\theta_{21}^{11}=\theta_{2}^{1}$ and $\theta_{11}^{21}=\theta_{12}^{22}=\theta_{1}^{2}$. In this case, the equilibrium ratio between the mean wealths of the two populations can be computed explicitly as
\begin{equation}
m_v^{2,\infty}=\xi m_v^{1,\infty}\qquad\text{with }\,\,\xi=\dfrac{\theta_{12}^{12}\gamma_1+\theta_{1}^{2}+\theta_{2}^{1}}{\theta_{21}^{21}\gamma_2+\theta_{1}^{2}+\theta_{2}^{1}}\,.
\end{equation}
If, instead, $\bar{\eta}\ne 0$, the system remains closed, but it no longer converges to an admissible equilibrium distribution. Instead, the mean wealths keep increasing, at a slow rate proportional to the small positive parameter $\bar{\eta}$. 

Finally, we turn to the dynamics of the total knowledge $M_x^i(t)$ in each country, obtained by testing the weak form \eqref{gen_kin_eq_inter} with $\varphi(x,v)=x$. The resulting macroscopic balance laws are
\begin{equation}\label{macro_Mx}
    \begin{sistem}
    \begin{aligned}
        \dfrac{d}{dt}M_x^1(t)=&M_x^2\left(\rho_1(t)\theta_{21}^{11}+\rho_2(t)\theta_{22}^{12}\right)-M_x^1\left(\rho_1(t)\theta_{11}^{21}+\rho_2(t)\theta_{12}^{22}\right)+\\[0.3cm]
        &\left(\bar{B}\rho_2^\lambda(t)-M_2^\lambda+M_2^\beta\right)\left(\theta_{21}^{11}\rho_1(t)+\theta_{22}^{12}\rho_2(t)\right)+\\[0.3cm]
        &\left(\bar{B}\rho_1^\lambda(t)-M_1^\lambda+M_1^\beta\right)\left(\theta_{11}^{11}\rho_1(t)+\theta_{12}^{12}\rho_2(t)+\theta_{11}^{12}\rho_1(t)+\theta_{12}^{11}\rho_2(t)\right)\\[0.5cm]
        \end{aligned}\\[1.5cm]
            \begin{aligned}
        \dfrac{d}{dt}M_x^2(t)=&M_x^1\left(\rho_1(t)\theta_{11}^{21}+\rho_2(t)\theta_{12}^{22}\right)-M_x^2\left(\rho_1(t)\theta_{21}^{11}+\rho_2(t)\theta_{22}^{12}\right)+\\[0.3cm]
        &\left(\bar{B}\rho_1^\lambda(t)-M_1^\lambda+M_1^\beta\right)\left(\theta_{11}^{21}\rho_1(t)+\theta_{12}^{22}\rho_2(t)\right)+\\[0.3cm]
        &\left(\bar{B}\rho_2^\lambda(t)-M_2^\lambda+M_2^\beta\right)\left(\theta_{22}^{22}\rho_2(t)+\theta_{21}^{21}\rho_1(t)+\theta_{21}^{22}\rho_1(t)+\theta_{22}^{21}\rho_2(t)\right)\,.\\
        \end{aligned}\\
    \end{sistem}
\end{equation}
Here, we use the notation:
\begin{equation}
M_i^\lambda(t):=\int\limits_{\mathbb{R}^2_+}x\lambda_i(x)f_i(t,x,v)dxdv\,,\qquad M_i^\beta(t):=\int\limits_{\mathbb{R}^2_+}x\beta_i(v)f_i(t,x,v)dxdv\,,
\end{equation}
and
\begin{equation}
\rho_i^\lambda:=\int\limits_{\mathbb{R}^2_+}\lambda^i_b(x)f_i(t,x,v)dxdv\,.
\end{equation}
These equations show that the evolution of knowledge in each country is governed by two distinct mechanisms. The first relates to migration and is captured by the initial terms in each equation, which describe the inflow and outflow of knowledge due to individuals moving between countries, redistributing knowledge proportionally to the migrant flows. The second mechanism concerns internal knowledge dynamics within each population, including learning, forgetting, and innovation, which drive the acquisition, retention, and depletion of knowledge over time. Overall, system \eqref{macro_Mx} illustrates that while migration primarily redistributes knowledge across countries, the balance between internal acquisition and loss ultimately determines whether the total knowledge in each population grows, stabilizes, or declines.

By combining \eqref{macro_rho} and \eqref{macro_Mx}, we obtain the evolution equations for the mean knowledge $m_x^i:=M_x^i/\rho_i$ of each country:
\begin{equation}\label{macro_mx}
    \begin{sistem}
    \begin{aligned}
        \dfrac{d}{dt}m_x^1(t)=&\dfrac{\rho_2(t)}{\rho_1(t)}\left(\rho_1(t)\theta_{21}^{11}+\rho_2(t)\theta_{22}^{12}\right)\left(m_x^2-m_x^1+\bar{B}\dfrac{\rho_2^\lambda(t)}{\rho_2(t)}-m_2^\lambda+m_2^\beta\right)+\\[0.3cm]
        &\left(\bar{B}\dfrac{\rho_1^\lambda(t)}{\rho_1(t)}-m_1^\lambda+m_1^\beta\right)\left(\theta_{11}^{11}\rho_1(t)+\theta_{12}^{12}\rho_2(t)+\theta_{11}^{12}\rho_1(t)+\theta_{12}^{11}\rho_2(t)\right)\\[0.5cm]
        \end{aligned}\\
       \begin{aligned}
        \dfrac{d}{dt}m_x^2(t)=&\dfrac{\rho_1(t)}{\rho_2(t)}\left(\rho_1(t)\theta_{11}^{21}+\rho_2(t)\theta_{12}^{22}\right)\left(m_x^1-m_x^2+\bar{B}\dfrac{\rho_1^\lambda(t)}{\rho_1(t)}-m_1^\lambda+m_1^\beta\right)+\\[0.3cm]
        &\left(\bar{B}\dfrac{\rho_2^\lambda(t)}{\rho_2(t)}-m_2^\lambda+m_2^\beta\right)\left(\theta_{22}^{22}\rho_2(t)+\theta_{21}^{21}\rho_1(t)+\theta_{21}^{22}\rho_1(t)+\theta_{22}^{21}\rho_2(t)\right)\,.\\[0.5cm]
        \end{aligned}
    \end{sistem}
\end{equation}
Here, $m_i^\lambda(t):=M_i^\lambda(t)/\rho_i(t)$ and $m_i^\beta(t):=M_i^\beta(t)/\rho_i(t)$. Similar to \eqref{macro_mv}, the system \eqref{macro_mx} is not closed, since its evolution equations involve additional moments of the distribution functions $f_i(t,x,v)$, for $i=1,2$. These additional quantities, namely $m_i^\lambda(t)$, $m_i^\beta(t)$, and $\rho^\lambda_i$ capture features of the knowledge distribution beyond the mean. The resulting equations describe how the mean knowledge of each population evolves under the joint influence of redistribution effects due to migration flows between the two countries and the endogenous acquisition or loss of knowledge within each country. 

Let assume $\theta_{22}^{12}=\theta_{21}^{11}=\theta_{2}^{1}$ and $\theta_{11}^{21}=\theta_{12}^{22}=\theta_{1}^{2}$, and set $\lambda_i(x)=\bar{\lambda}$, $\lambda^i_b(x)=\bar{\lambda}_b$, and $\beta_i(v)=\bar{\beta}$, for $i=1,..,n$, with $\bar{\lambda}>\bar{\beta}$. Further setting $\bar{\lambda}_b=1$, for simplicity, system \eqref{macro_mx} reduces to a closed set of equations for the mean knowledge $m_x^1(t)$ and $m_x^2(t)$:
\begin{equation}\label{macro_mx_red}
    \begin{sistem}
    \begin{aligned}
        \dfrac{d}{dt}m_x^1(t)=&\dfrac{\rho_2(t)}{\rho_1(t)}\theta_{2}^{1}\bar{\rho}\left((1+\bar{\beta}-\bar{\lambda})m_x^2-m_x^1+\bar{B}\right)+\\[0.3cm]
        &\left(\bar{B}+(\bar{\beta}-\bar{\lambda})m_x^1\right)\left(\chi_{11}\rho_1(t)+\chi_{12}\rho_2(t)+\theta_{1}^{2}\rho_1(t)+\theta_{2}^{1}\rho_2(t)\right)\\[0.5cm]
        \end{aligned}\\[0.5cm]
       \begin{aligned}
        \dfrac{d}{dt}m_x^2(t)=&\dfrac{\rho_1(t)}{\rho_2(t)}\theta_{1}^{2}\bar{\rho}\left((1+\bar{\beta}-\bar{\lambda})m_x^1-m_x^2+\bar{B}\right)+\\[0.3cm]
        &\left(\bar{B}+(\bar{\beta}-\bar{\lambda})m_x^2\right)\left(\chi_{22}\rho_2(t)+\chi_{21}\rho_1(t)+\theta_{1}^{2}\rho_1(t)+\theta_{2}^{1}\rho_2(t)\right)\,.\\[0.5cm]
        \end{aligned}
    \end{sistem}
\end{equation}
To study the equilibrium distributions of the system, we rewrite it in matrix form as
\begin{equation}\label{system_eq_disti}
\mathbf{A} \underline{{\bf m}}=\underline{{\bf b}}
\end{equation}
where $\underline{{\bf m}}:=(m_x^1, m_x^2)^T$ denotes the vector of unknowns. Bearing in mind that, under these assumptions, from \eqref{alpha} we get 
$\rho_2^{\infty}=\alpha\rho_1^{\infty}$  
with $\alpha=\theta_{1}^{2}/\theta_{2}^{1}$ and the quantity $\bar{\rho}=\rho_1+\rho_2$ is preserved and, for simplicity, normalized to $\bar{\rho}=1$, the coefficient matrix $\mathbf{A}$ is given by
\begin{equation}\label{matrix_A}
\begin{split}
&\mathbf{A}:=
\begin{pmatrix}
        a_{11}& a_{12} \\
        a_{21}& a_{22}
    \end{pmatrix}=\\[0.5cm]
    &\bar{\rho}
    \begin{pmatrix}
        -\theta_1^2-\dfrac{\bar{\lambda}-\bar{\beta}}{\theta_1^2+\theta_2^1}\left(\chi_{11}\theta_2^1+\chi_{12}\theta_1^2+2\theta_1^2\theta_2^1\right) & \theta_1^2(1+\bar{\beta}-\bar{\lambda})\\[0.4cm]
        \theta_2^1(1+\bar{\beta}-\bar{\lambda}) &-\theta_2^1-\dfrac{\bar{\lambda}-\bar{\beta}}{\theta_1^2+\theta_2^1}\left(\chi_{22}\theta_1^2+\chi_{12}\theta_2^1+2\theta_1^2\theta_2^1\right)
    \end{pmatrix}\,,
    \end{split}
\end{equation}
and the right-hand side vector $\underline{{\bf b}}$ reads
\begin{equation}
\underline{{\bf b}}:=
\begin{pmatrix}
        b_{1} \\
        b_{2}
    \end{pmatrix}=\bar{\rho}\bar{B}
    \begin{pmatrix}
        \theta_1^2+\dfrac{1}{\theta_1^2+\theta_2^1}\left(\chi_{11}\theta_2^1+\chi_{12}\theta_1^2+2\theta_1^2\theta_2^1\right)\\[0.4cm]
        \theta_2^1+\dfrac{1}{\theta_1^2+\theta_2^1}\left(\chi_{22}\theta_1^2+\chi_{12}\theta_2^1+2\theta_1^2\theta_2^1\right)
    \end{pmatrix}\,.
\end{equation}
It can be shown that $\det(\mathbf{A}) \neq 0$, and therefore system \eqref{system_eq_disti} admits a unique solution, $$\underline{{\bf m}}^{\infty}:=(m_x^{1,\infty}, m_x^{2,\infty})^T\,,$$ which represents the equilibrium mean knowledge levels of the two nations. In particular, if we drop the assumption $\bar{\lambda} > \bar{\beta}$ and, for simplicity, take $\bar{\lambda} = \bar{\beta} = 1$, the matrix \eqref{matrix_A} reduces to
\begin{equation}
\mathbf{A}=\bar{\rho}
    \begin{pmatrix}
        -\theta_1^2 & \theta_1^2\\[0.4cm]
        \theta_2^1 &-\theta_2^1
    \end{pmatrix}\,,
\end{equation}
for which $\det(\mathbf{A}) = 0$. However, since the determinant $\det([\mathbf{A}\,|\,\underline{\mathbf{b}}]) \neq 0$, system \eqref{macro_mx_red} does not admit an equilibrium distribution. This result is consistent with the fact that the condition $\bar{\lambda} > \bar{\beta}$ is required in the microscopic model of knowledge evolution to ensure the existence of an upper bound for the mean knowledge, as discussed in Section~\ref{national_trade}.

\subsubsection{Quasi-invariant limit}
Considering the Fokker–Planck system of equations describing international markets with individual transfers among $n$ different countries, as given in \eqref{FP_interna}, we now specialize this general result to the case of the  two-country economy introduced in Section \ref{2C_economy}. Our main goal is to derive the Fokker–Planck system consisting of two equations for the distributions $g_1(\tau,x,v)$ and $g_2(\tau,x,v)$, and to discuss some aspects of their (partial) asymptotic behavior. To this end, we consider a simplified setting in which the probability of transfer to the $i$-th subgroup does not depend on the countries of the interacting agents. In particular, we assume
$\theta_{2}^{1}:=\theta_{22}^{12}=\theta_{21}^{11}$ and $\theta_{1}^{2}:=\theta_{11}^{21}=\theta_{12}^{22}$, so that $
\rho_2^{\infty}=\alpha\rho_1^{\infty}
$
with $\alpha=\theta_{1}^{2}/\theta_{2}^{1}$ and, since $\bar{\rho}=\rho_1+\rho_2 = 1$,
\begin{equation}
    \rho_1^{\infty}=\dfrac{1}{1+\alpha}\qquad\text{and}\qquad \rho_2^{\infty}=\dfrac{\alpha}{1+\alpha}\,.
\end{equation}
Expanding the distribution functions $g_1$ and $g_2$ in powers of $\varepsilon$ while enforcing that the globally conserved quantities (i.e. the zeroth moment) remain unexpanded leads to the first-order constraint
\begin{equation}
    \rho_1^{(1)}+ \rho_2^{(1)}=0\,.
\end{equation}
Proceeding exactly as in Section \ref{2C_economy}, from \eqref{g0_s_QI} one obtains the leading-order relation between the two distributions
\begin{equation}\label{g10_rel}
    g_1^{(0)}=\dfrac{1}{\alpha}g_2^{(0)} 
\end{equation}
with $\alpha$ defined as above. Throughout we assume identical stochastic fluctuation amplitudes across different interaction types, namely $\delta_i=\delta$ and $\sigma_{ik}=\sigma$ for ${i,k=1,2}$. We also observed that $\rho_2^{(0)}=\alpha \rho_1^{(0)}$, i.e, $\rho_i^{(0)}=\rho_i^{\infty}$ for $i=1,2$. For $s=1$, the Fokker–Planck equation \eqref{FP_interna} reduces to 
\begin{equation}\label{g1_QI}
\begin{split}
    &\dfrac{d}{d\tau} g_1^{(0)}=-\dfrac{\partial}{\partial x} \left(g_1^{(0)}\left(\mathcal{A}^1_x\dfrac{\bar{K}_1}{1+\alpha}+\theta_{1}^{2}\mathcal{A}^2_x\right)\right)-\\[0.3cm]
    &\dfrac{\partial}{\partial v} \left(g_1^{(0)}\left(\mathcal{A}^{11}_v(\chi_{11}+\theta_{1}^{2})+\mathcal{A}^{12}_v(\chi_{12}+\theta_{2}^{1})+ \theta_{1}^{2}\left(\mathcal{A}^{21}_v+\mathcal{A}^{22}_v\right)\right)\right)+\\[0.3cm]
    &\dfrac{\delta^2}{2(1+\alpha)}\dfrac{\partial^2}{\partial x^2} \left( x^2 g_1^{(0)}\left(\bar{K}_1 +\theta_{1}^{2} (1+\alpha)\right)\right)+\dfrac{\sigma^2}{2(1+\alpha)}\dfrac{\partial^2}{\partial v^2} \left(v^2 g_1^{(0)}\left(\bar{K}_1 +\theta_{1}^{2} (1+\alpha)\right)\right)+\\[0.3cm]
    &\left(\theta_{2}^{1}g_2^{(1)}-\theta_{1}^{2}g_1^{(1)}\right)\,,
     \end{split}
\end{equation}
where we define $\bar{K}_1:=(\chi_{11}+\theta_{1}^{2})+\alpha(\chi_{12}+\theta_{2}^{1})$. Analogously, for $s=2$, the equation for $g_2^{(0)}$ reads
\begin{equation}\label{g2_QI}
\begin{split}
    &\dfrac{d}{d\tau} g_2^{(0)}=-\dfrac{\partial}{\partial x} \left(g_2^{(0)}\left(\mathcal{A}^2_x\dfrac{\bar{K}_2}{1+\alpha}+\theta_{2}^{1}\mathcal{A}^1_x\right)\right)-\\[0.3cm]
    &\dfrac{\partial}{\partial v} \left(g_2^{(0)}\left(\mathcal{A}^{22}_v(\chi_{22}+\theta_{2}^{1})+\mathcal{A}^{21}_v(\chi_{21}+\theta_{1}^2)+ \theta_{2}^{1}\left(\mathcal{A}^{12}_v+\mathcal{A}^{11}_v\right)\right)\right)+\\[0.3cm]
    &\dfrac{\delta^2}{2(1+\alpha)}\dfrac{\partial^2}{\partial x^2} \left( x^2 g_2^{(0)}\left(\bar{K}_2 +\theta_{2}^{1} (1+\alpha)\right)\right)+\dfrac{\sigma^2}{2(1+\alpha)}\dfrac{\partial^2}{\partial v^2} \left(v^2 g_2^{(0)}\left(\bar{K}_2 +\theta_{2}^{1} (1+\alpha)\right)\right)+\\[0.3cm]
    &\left(\theta_{1}^{2}g_1^{(1)}-\theta_{2}^{1}g_2^{(1)}\right)\,,
     \end{split}
\end{equation}
with $\bar{K}_2:=\alpha(\chi_{22}+\theta_{2}^{1})+(\chi_{21}+\theta_{1}^{2})$. Equations \eqref{g1_QI} and \eqref{g2_QI} are Fokker–Planck equations with additional reaction terms. The first three lines on the right-hand side describe the drift and diffusion contributions, whose effective coefficients (drift velocity and diffusion strength) depend on the market parameters, the knowledge dynamics, and the interaction frequencies. The last line represents a reaction term involving only the first-order corrections $g_i^{(1)}$. This term encodes the interspecies exchange mechanism, which redistributes mass between the two populations without altering the global conservation law. 

By summing equations \eqref{g1_QI} and \eqref{g2_QI}, we obtain a Fokker–Planck equation for the total distribution $g_1^{(0)}+g_2^{(0)}$ without a reaction term. Using relation \eqref{g10_rel}, this equation can be equivalently expressed as
\begin{equation}\label{g1sumg2_QI}
\begin{split}
    (1+\alpha)\dfrac{d}{d\tau} g_1^{(0)}=&-\dfrac{\partial}{\partial x} \left[g_1^{(0)}\left(\mathcal{A}^1_x\left(\dfrac{\bar{K}_1}{1+\alpha}+\alpha\theta_{2}^{1}\right)+\mathcal{A}^2_x\left(\dfrac{\alpha \bar{K}_2}{1+\alpha}+\theta_{1}^{2}\right)\right)\right]\\[0.3cm]
    &-\dfrac{\partial}{\partial v} \left[g_1^{(0)}\left[\mathcal{A}^{11}_v\left(\chi_{11}+2\theta_{1}^{2}\right)+\alpha\mathcal{A}^{22}_v\left(\chi_{22}+2\theta_{2}^{1}\right)\right]\right]\\[0.2cm]
    &-\dfrac{\partial}{\partial v} \left[g_1^{(0)}\left[\mathcal{A}^{12}_v\left(\chi_{12}+(\theta_{2}^{1}+\theta_{1}^{2})\right)+\alpha\mathcal{A}^{21}_v\left(\chi_{21}+(\theta_{2}^{1}+\theta_{1}^{2})\right)\right]\right]\\[0.3cm]
    &+\dfrac{\delta^2}{2(1+\alpha)}\dfrac{\partial^2}{\partial x^2} \left[x^2 g_1^{(0)}\left(\bar{K}_1+\alpha\bar{K}_2 +2\theta_{1}^{2} (1+\alpha)\right)\right]\\[0.2cm]
    &+\dfrac{\sigma^2}{2(1+\alpha)}\dfrac{\partial^2}{\partial v^2} \left[v^2 g_1^{(0)}\left(\bar{K}_1+\alpha\bar{K}_2 +2\theta_{1}^{2} (1+\alpha)\right)\right]\,.
     \end{split}
\end{equation}
For readability, in equations \eqref{g1_QI}, \eqref{g2_QI}, and \eqref{g1sumg2_QI} we neglect the dependence on the set of variables $(\tau,x,v)$. 

Considering equation \eqref{g1sumg2_QI} and focusing on the partial steady state in the wealth variable $v$, we follow an approach analogous to Section \ref{QI_natio}. Specifically, assuming that $\phi_i(x,v)$ is linearly increasing with respect to $v$, i.e., $\phi_i(x,v)=\tilde{\phi}_i(x)v$, for $i=1,2$, we obtain 
\begin{equation}\label{partial_v_s1}
g_{1,v}^{\infty}(x,v)=\mathcal{C}_{1,v}\, v^{-2\left(\dfrac{\mathcal{I}_1^v}{2}+1\right)} \,\,\,\,e^{-\dfrac{\nu}{v}\Lambda(M_1^{\psi},M_2^{\psi})}\,,
\end{equation}
where $\mathcal{C}_{1,v}$ is a normalization constant. Here,
\begin{equation}
    \nu:=\dfrac{2(1+\alpha)}{\sigma^2(\bar{K}_1+\alpha\bar{K}_2+2\theta_1^2(1+\alpha))}
\end{equation}
and
\begin{equation}
\begin{split}
    \Lambda(M_1^{\psi},M_2^{\psi}):=&\,\gamma_1 M_1^\psi\left(\bar{K}_1+\theta_1^2(1+\alpha)+\alpha\left(\chi_{21}-\chi_{12}\right)\right)+\\[0.2cm]
    &\dfrac{\gamma_2}{\alpha}M_2^\psi\left(\alpha\bar{K}_2+\theta_1^2(1+\alpha)+\alpha\left(\chi_{12}-\chi_{21}\right)\right)\,,
    \end{split}
\end{equation}
while 
\begin{equation}
\begin{split}
\mathcal{I}_1^v:=&\,\dfrac{\nu}{(1+\alpha)} \Bigg[\left(\gamma_1\psi_1(x)-\bar{\eta}\tilde{\phi}_1(x)\right)\left(\bar{K}_1+\theta_1^2(1+\alpha)\right)\Bigg]+\\[0.3cm]
&\dfrac{\nu}{(1+\alpha)} \Bigg[\left(\gamma_2\psi_2(x)-\bar{\eta}\tilde{\phi}_2(x)\right)\left(\alpha\bar{K}_2+\theta_1^2(1+\alpha)\right)\Bigg]\,.
\end{split}
\end{equation}
Since $\bar{\eta}$ is assumed sufficiently small, we can consider that the exponent $\mathcal{I}_1^v/2+1$ in \eqref{partial_v_s1} is always positive. Consequently, $g_{1,v}^{\infty}(x,v)\to0$ as $v\to0^+$, while for $v\to+\infty$ it decays as an inverse power law $v^{-(1+I_1^v)}$, where $I_1^v$ is the Pareto index,\cite{gualandi2018pareto} given by
\begin{equation}
\begin{split}
&I_1^v:\,=\mathcal{I}_1^v+1=\\[0.3cm]
&\dfrac{2}{\sigma^2(\bar{K}_1+\alpha\bar{K}_2+2\theta_1^2(1+\alpha))}\Bigg[\left(\gamma_1\psi_1(x)-\bar{\eta}\tilde{\phi}_1(x)\right)\left(\bar{K}_1+\theta_1^2(1+\alpha)\right)\Bigg]+\\[0.3cm]
&\dfrac{2}{\sigma^2(\bar{K}_1+\alpha\bar{K}_2+2\theta_1^2(1+\alpha))}\Bigg[\left(\gamma_2\psi_2(x)-\bar{\eta}\tilde{\phi}_2(x)\right)\left(\alpha\bar{K}_2+\theta_1^2(1+\alpha)\right)\Bigg]\!\!+\!1\,.
\end{split}
\end{equation}
The Pareto index depends on the trading and saving propensities of both populations, described by $\psi_i$, $\tilde{\phi}_i$, and $\gamma_i$, on the interaction rates $\theta_{ik}^{i'k'}$ (directly or through $\alpha$), and on the stochastic parameter $\sigma^2$, with $i,k,i',k'\in\{1,2\}$. Since, according to \eqref{g10_rel}, $g_2^{(0)}=\alpha g_1^{(0)}$, both populations share the same Pareto index.

We notice that, the case of national trade, where only a single population is present and, thus, no individual transfers occur, corresponds to setting $\gamma_1=\gamma_2$, $\psi_1(x)=\psi_2(x)$, $\tilde{\phi}_1(x)=\tilde{\phi}_2(x)$, $\chi_{11}=\chi_{22}=\chi_{12}=\chi_{21}=\chi$, as well as ${\theta_1^2=\theta_2^1=\theta}$. In this scenario, we have $\alpha=1$, so that $g_1^{(0)}= g_2^{(0)}$, and moreover ${\bar{K}_1=\bar{K}_2=2\chi+2\theta}$. Consequently, the Pareto index simplifies to
\begin{equation}\label{PIred_1D}
\begin{split}
&I_1^v=\dfrac{1}{2\sigma^2(\chi+2\theta)}\Bigg[4\left(\gamma\psi(x)-\bar{\eta}\tilde{\phi}(x)\right)(\chi+2\theta)\Bigg]+1= \dfrac{2}{\sigma^2}\left[\gamma\psi(x)-\bar{\eta}\tilde{\phi}(x)\right]+1
\end{split}
\end{equation}
which coincides exactly with the expression obtained in Section~\ref{QI_natio}. Furthermore, if we assume that the saving and trading propensity functions of the two populations are identical, then even when the interaction rates differ, i.e. $\theta_1^2 \ne \theta_2^1$, the Pareto index of each subroup still coincides with that of the single-population case \eqref{PIred_1D}.

\section{Conclusion and perspectives}
In this work, we propose a kinetic model to describe the dynamical evolution of wealth and knowledge distributions in both national and international markets, explicitly accounting for the possibility of individuals transferring across countries. 

Starting from the scenario of a domestic market, we define the microscopic interaction rules between two agents within the same country that govern changes in their wealth and knowledge. From these rules, we derive the corresponding Boltzmann equation in weak form. Our approach follows the framework introduced in Refs.~\cite{cordier2005kinetic,pareschi2014wealth}, which has already been shown to reproduce realistic long-term behaviors. The main novelties of our model are twofold. First, we introduce an explicit coupling between wealth and knowledge: the evolution of an agent’s knowledge depends on its wealth, while the dynamics of wealth are simultaneously influenced by the agent’s knowledge. Second, the impact of knowledge on risk-related market fluctuations is modeled through a random variable with nonzero mean, capturing the effects of an open economy. This mechanism reflects the role of investments in sustaining long-term economic growth, leading to a slow increase in the total wealth of the system, in accordance with historical observations. By analyzing the evolution of the mean knowledge and wealth distributions, under suitable assumptions, we prove that the mean knowledge remains bounded, while the mean wealth exhibits exponential growth. We then apply quasi-invariant limit techniques to investigate the asymptotic behavior of the model in the regime of small interactions. In this limit, we recover a Fokker–Planck type equation for the agent distribution, which highlights the coupling between wealth and knowledge through the emergent drift terms. We study the partial steady states with respect to knowledge and wealth separately, showing that in the long-time behavior both distributions develop Pareto tails, in agreement with realistic economic scenarios.\cite{cordier2005kinetic,gualandi2018pareto}

To investigate the dynamics of a global market, we extend the framework by incorporating an additional microscopic variable. This variable labels each agent according to the country of origin, as well as the probability of individual transfers between countries during interactions. Within this setting, we derive the weak formulation of the corresponding Boltzmann equations governing the evolution of the agent distribution in each subgroup. Moreover, in the quasi-invariant interaction limit, we recover the associated Fokker–Planck system. Unlike the purely national trade scenario, the resulting Fokker–Planck equations include reaction terms directly related to inter-country transfers, which induce changes in the population size of each subgroup. This effect becomes even more evident when analyzing the evolution of the moments of the distribution function. In particular, considering a two-country economy with domestic and international trade (both without migration and with single-agent migration) we derive macroscopic equations for the evolution of the population size, as well as the mean wealth and mean knowledge within each subgroup. The population size evolves solely due to transfers, with its dynamics depending explicitly on the interaction probabilities. By contrast, it is generally not possible to obtain a closed macroscopic system for the mean wealth and knowledge because of their intrinsic coupling. However, under suitable simplifying assumptions on the coefficients involved, one can prove the existence of equilibrium distributions for the mean knowledge of the two populations, together with the exponential growth of the mean wealth. Finally, by specializing the quasi-invariant limit results to the case of two-country markets, we analyze the partial steady-state distribution of wealth and derive an explicit expression for the Pareto index. This index directly reflects the trading and saving propensities of both populations, as well as the interaction and transfer probabilities. 

As future work, the kinetic description developed in this paper will be complemented by numerical experiments. These will aim at investigating and comparing the evolution of both the Boltzmann-type equations and their Fokker–Planck approximations, as well as the dynamics of the associated macroscopic quantities. They will not only validate the theoretical findings but also provide deeper insight into transient behaviors, the role of migration and transfers, and the emergence of long-term equilibria under different market configurations. Moreover, since our asymptotic analysis indicates that distributions with power-law tails also emerge in the knowledge distribution within the populations, it would be interesting to explore this phenomenon in greater detail. In particular, understanding the mechanisms behind the formation of these tails and their impact on the evolution of national and global markets would be desirable. Finally, a natural extension of this work would be to compare the model with empirical data, evaluating its relevance for real-world market dynamics.

\section*{Acknowledgment}
The research of all authors has been carried out under the auspices of GNFM (National Group of Mathematical-Physics) of INDAM (National Institute of Advanced Mathematics). MC was supported by the European Union and the Italian Ministry of University and Research through the PNRR project Young Researchers 2024-SOE {\it ‘Integrated Mathematical Approach to Tumor Interface Dynamics’} (CUP: E13C24002380006). MC has been partially supported by the State Research Agency of the Spanish Ministry of Science and FEDER-EU, project PID2022-137228OB-I00 (MICIU/AEI /10.13039/501100011033); by Modeling Nature Research Unit, Grant QUAL21-011 funded by Consejería de Universidad, Investigaci\'on e Innovaci\'on (Junta de Andalucía). The work has been performed in the frame of the project PRIN 2022 PNRR {\it ‘Mathematical Modelling for a Sustainable Circular Economy in Ecosystems’} (project code P2022PSMT7, CUP: D53D23018960001) funded by the European Union - NextGenerationEU and by MUR-Italian Ministry of Universities and Research. 


\end{document}